## Statistics of Residual Stress in Random Microstructures: Mean-Field Estimates and Full-Field Validations

Tarkes Dora Pallicity

Microsolidics Group, Department of Mechanical Engineering, Indian Institute of Technology Guwahati, India
tarkes.pallicity@iitg.ac.in

### Abstract

Fluctuations of local fields are crucial for prediction of failure in random composites across different scales as well as estimating the inelastic behavior of it. This can be quantified statistically through second moments of the local fields which can be quickly estimated using mean-field homogenization (MFH). However, the exact fluctuation field can be estimated using full-field methods though it comes at the cost of intensive computational resources and limited scalability to complex microstructures. In this work, MFH is used to estimate the statistical variation of the field quantities and then cross-verified with full-field methods for a linear-thermoelastic homogenization problem. Analytical expression to calculate the second moments of the local fields for a linear thermo-elastic problem using MFH is obtained based on Hill-Mandel condition. The expressions fundamentally rely on the solution of linear elastic problem which in turn depends on the derivatives of Hill's polarization tensor. Solution of this derivative term has been analytically and semi-analytically derived in previous work [1]. The statistical distribution of residual stress tensor components and equivalent stress in particulate and unidirectional fibrous composites, arising purely due to differential thermal expansion, is computed and compared with full-field homogenization. Full-field simulations indicated non-Gaussian distribution of stress components whereas Weibull-like distributions for equivalent residual stress. Nevertheless, the assumed Gaussian distribution in mean-field estimates captures the essential features.

## 1 Introduction

Mean-field homogenization method based on Eshelby's solution have been conventionally used to estimate the first statistical moments (mean) of the field quantities and effective properties. It has been established that such methods not only provide mean but also statistics of the local field quantities via full second statistical moments of the field quantities. In a small strain framework, a full second-order statistical moment is characterized by a positive definite, minor- and major-symmetric fourth-order tensor. Field fluctuations have been incorporated in second-order homogenization methods via specific scalar projections of the second moments of the local field to make estimates of the effective nonlinear behavior of heterogenous materials [2–5]. Full second moments of stress field can be useful in estimating the fluctuation of field invariants to model the localized failure mechanisms.

Expressions to compute full second-moments of local fields in linear problems was first established by Bobeth and Diener [6, 7]. Later, Kreher and Pompe [8] provided the strategy to derive exact expressions to calculate second statistical moments of field quantities in heterogeneous thermoelastic materials. Semi-analytical strategy to compute full second moments for linear composites and geometrically non-linear composites is derived in [9, 10] and [11] respectively. A complete analytical solution was derived for calculating second moments in heterogenous materials comprising of isotropic linear elastic phases with varying aspect ratio of ellipsoidal inhomogeneities [1].

Full-field homogenization methods based on finite element methods (FE) provide exact statistical distribution of tensorial field quantities. Statistical comparisons have been done between the full-field solutions and corresponding analytical solutions, both at the level of individual tensor components and their invariants [1]. However, such comparisons are lacking for linear thermoelastic problems. In this work, the gap is addressed by examining the statistical distribution of residual stresses that develop during the cooling of composites. This case is chosen for its practical importance, as it is representative of process-induced stresses [12] and thermo-mechanical service loading conditions[13].

In this paper Section 2 describes the constitutive behavior of the phases in the composites. Section 3 establishes the generalized relations for computing first and second moments of local fields of a linear thermoelastic homogenization problem consisting of $n$ phases. Later the relations for second moment calculation are simplified for 2-phase composites with random microstructures. The derivatives of the Hill's polarization tensor that is required for computing second moments is elaborated in Section 4. Sampling of the random field data considering normal distribution of local fields is discussed in 5. Section 6 establishes the framework for full-field simulations using FE method. Simulations are carried out to compute exact statistical distribution of local residual stress fields generated in the class of random microstructures due to cooling. The results from both the mean-field and full-field solutions are compared for the distribution of equivalent residual stress as well as the components of the residual stress tensor. The comparison is carried out for composites with ellipsoidal inhomogeneities of aspect ratio unity and infinity.

### 1.1 Nomenclature

A symbolic tensor notation is used throughout the text. The scalar quantities are denoted by light-face type Latin and Greek characters e.g., $w, K, G$. An orthonormal basis $\{\boldsymbol{e}_1, \boldsymbol{e}_2, \boldsymbol{e}_3\} = \{\boldsymbol{e}_i\}$ of a three-dimensional Euclidean space is used throughout to define the tensorial quantities. The first-order tensors are denoted by bold lower-case Latin characters e.g., $\boldsymbol{x}$. The second-order tensors are denoted by bold upper-case Latin characters and Greek characters e.g., $\boldsymbol{I}, \boldsymbol{\Sigma}, \boldsymbol{E}, \boldsymbol{\sigma}, \boldsymbol{\varepsilon}$. The fourth-order tensors are denoted by blackboard bold upper-case Latin characters e.g., $\mathbb{C}, \mathbb{I}$. Similarly, the eighth order tensor is represented by calligraphic bold upper-case English alphabets e.g. $\boldsymbol{\mathcal{I}}$. The major transpose operation is indicated as $(\bullet)^{\mathrm{T_H}}$, left and right minor symmetry is indicated as $(\bullet)^{\mathrm{T_L}}$ and $(\bullet)^{\mathrm{T_R}}$ respectively for fourth order tensors. In case of higher order tensors $(\bullet)^{\mathrm{T_H}}{}_{i_1\cdots i_n i_{n+1}\cdots i_{2n}} = (\bullet)_{i_{n+1}\cdots i_{2n} i_1\cdots i_n}$ , $(\bullet)^{\mathrm{T_L}}{}_{i_1 i_2 i_3\cdots i_n} = (\bullet)_{i_3\cdots i_n i_1 i_2}$ and $(\bullet)^{\mathrm{T_R}}{}_{i_1\cdots i_{n-2} i_{n-1} i_n} = (\bullet)_{i_{n-1} i_n i_1\cdots i_{n-2}}$. The linear transformation of a vector by a second order tensor is denoted by $\boldsymbol{A}\boldsymbol{x} = A_{ij} x_j \boldsymbol{e}_i$, while higher-order linear maps are indicated by $\mathbb{C}[\boldsymbol{\varepsilon}] = C_{ijkl}\varepsilon_{kl}(\boldsymbol{e}_i \otimes \boldsymbol{e}_j)$. The scalar product between tensors of same order is denoted as $\big((\blacksquare)\cdot(\bullet)\big) = (\blacksquare)_{i_1\cdots i_n}(\bullet)_{i_1\cdots i_n}$. The tensor product of different orders of tensors are indicated as $\big((\blacksquare)\otimes(\bullet)\big)_{i_1\cdots i_m i_n\cdots i_o} = (\blacksquare)_{i_1\cdots i_m}(\bullet)_{i_n\cdots i_o}$. The Kronecker product [14] of any order tensors of is defined as $\big((\blacksquare)\times(\bullet)\big)_{abc\ldots ijk\ldots mno\ldots xyz\ldots} = (\blacksquare)_{abc\ldots mno\ldots}(\bullet)_{ijk\ldots xyz\ldots}$. The Frobenius norm of a tensor is denoted as for e.g. $\parallel \bullet \parallel$. The $n$-times tensor product and Kronecker product of any tensorial quantities with itself is symbolically indicated as $(\bullet)^{\otimes n}$ and $(\bullet)^{\times n}$ respectively. The Kronecker delta is indicated by $\delta_{ij}$. The second, fourth-order (both major and minor symmetric) and eighth order identity tensors is denoted by $\boldsymbol{I}$, $\mathbb{I}^{\mathrm{s}} = \frac{1}{2}(\boldsymbol{I}^{\times 2} + (\boldsymbol{I}^{\times 2})^{\mathrm{T_R}})$ and $\boldsymbol{\mathcal{I}}^{\mathrm{s}}$, respectively. The definition of $\boldsymbol{\mathcal{I}}^{\mathrm{s}}$ is $[\boldsymbol{\mathcal{I}}^{\mathrm{s}}]_{ijklmnop} = \frac{1}{8}(\delta_{im}\delta_{jn}\delta_{ko}\delta_{lp} + \delta_{jm}\delta_{in}\delta_{ko}\delta_{lp} + \delta_{im}\delta_{jn}\delta_{lo}\delta_{kp} + \delta_{jm}\delta_{in}\delta_{lo}\delta_{kp} + \delta_{km}\delta_{ln}\delta_{io}\delta_{jp} + \delta_{lm}\delta_{kn}\delta_{io}\delta_{jp} + \delta_{km}\delta_{ln}\delta_{jo}\delta_{ip} + \delta_{lm}\delta_{kn}\delta_{jo}\delta_{ip})$. The fourth-order orthogonal projection tensors are denoted and defined as $\mathbb{P}_1 = \frac{1}{3}(\boldsymbol{I}\otimes\boldsymbol{I})$ and $\mathbb{P}_2 = \mathbb{I}^{\mathrm{s}} - \mathbb{P}_1$ corresponding to the spherical and deviator, respectively. The fourth order positive definite major and minor symmetric elastic stiffness tensor is represented as $\mathbb{C}$. The spatial average of the field quantities over a $\gamma$-phase of a heterogeneous material is indicated by angular brackets for e.g., $\langle\boldsymbol{\sigma}\rangle_\gamma$.

## 2 Effective Constitutive Behavior

Consider the microstructure of a heterogenous material occupying a volume $\omega$, consisting of uniformly distributed inhomogeneities in a matrix (two phase). The geometric model of the random microstructure can be sampled by a representative volume element (RVE) under the assumptions of ergodicity and length scale separation. The microstructure has a specific realization where phase 1 is considered to be matrix. Each $\gamma$-phase is characterized by indicator function $\mathcal{I}_\gamma(\boldsymbol{x})$ [15] such that,

$$\mathcal{I}_\gamma(\boldsymbol{x}) = \begin{cases} 1 & \boldsymbol{x} \in \omega_\gamma \\ 0 & \text{else,} \end{cases}$$

where $\boldsymbol{x}$ is the local coordinates lying in $\omega_\gamma$ of the composite which occupies a volume fraction $c_\gamma$.

In a small strain framework, the constitutive relations for a linear thermoelastic phase in a heterogenous material is given as,

$$\begin{aligned} \boldsymbol{\varepsilon}(\boldsymbol{x}) &= \mathbb{S}[\boldsymbol{\sigma}(\boldsymbol{x})] + \boldsymbol{\varepsilon}_\theta, \\ \boldsymbol{\sigma}(\boldsymbol{x}) &= \mathbb{C}[\boldsymbol{\varepsilon}(\boldsymbol{x})] - \boldsymbol{\beta}(\boldsymbol{x}), \end{aligned} \tag{1}$$

where, $\mathbb{S}$ and $\mathbb{C}$ are the compliance and stiffness tensors respectively which are considered to be homogenous over each phase in the composite. The terms $\boldsymbol{\varepsilon}(\boldsymbol{x})$, $\boldsymbol{\varepsilon}_\theta$ and $\boldsymbol{\beta}(\boldsymbol{x})$ is the local total strain field, thermal strain (or stress-free strains) and thermal stress respectively. In the context of homogenization, the term $\boldsymbol{\varepsilon}_\theta$ is non-zero and homogenous because of temperature change is assumed to be uniform in all the phases of the microstructure. Since linear thermoelastic materials falls under the class of generalized standard materials, the strain energy density function (or stress potential) for a linear thermoelastic material can be identified to define the constitutive behavior of the phase. It is given as,

$$\begin{aligned} 2w(\boldsymbol{x}) &= (\boldsymbol{\varepsilon}(\boldsymbol{x}) - \boldsymbol{\varepsilon}_\theta) \cdot \mathbb{C}[\boldsymbol{\varepsilon}(\boldsymbol{x}) - \boldsymbol{\varepsilon}_\theta] \\ &= \mathbb{C} \cdot \Big(\boldsymbol{\varepsilon}(\boldsymbol{x})^{\otimes 2} + {\boldsymbol{\varepsilon}_\theta}^{\otimes 2} - (\boldsymbol{\varepsilon}(\boldsymbol{x}) \otimes \boldsymbol{\varepsilon}_\theta) - (\boldsymbol{\varepsilon}_\theta \otimes \boldsymbol{\varepsilon}(\boldsymbol{x}))\Big). \end{aligned} \tag{2}$$

The state of stress at each local position in the phase is given as,

$$\boldsymbol{\sigma}(\boldsymbol{x}) = \frac{\partial w(\boldsymbol{x})}{\partial \boldsymbol{\varepsilon}(\boldsymbol{x})}.$$

Alternatively, the complimentary strain energy function (or strain potential) of the phase can be defined,

$$u(\boldsymbol{x}) = \sup_{\boldsymbol{\varepsilon}} \big(\boldsymbol{\sigma}(\boldsymbol{x}) \cdot \boldsymbol{\varepsilon}(\boldsymbol{x}) - w(\boldsymbol{x})\big).$$

The state of the total strain is then given as,

$$\boldsymbol{\varepsilon}(\boldsymbol{x}) = \frac{\partial u(\boldsymbol{x})}{\partial \boldsymbol{\sigma}(\boldsymbol{x})}.$$

The strain potential $u(\boldsymbol{x}, \boldsymbol{\sigma})$ can be obtained by the Legendre-Fenchel transform of $w(\boldsymbol{x}, \boldsymbol{\varepsilon})$ and vice-a-versa due to convexity nature of the functions.

The overall behavior of the heterogenous materials due to temperature change is given as,

$$\begin{aligned} \boldsymbol{E} &= \overline{\mathbb{S}}[\boldsymbol{\Sigma}] + \boldsymbol{E}_\theta, \\ \boldsymbol{\Sigma} &= \overline{\mathbb{C}}[\boldsymbol{E}] - \boldsymbol{B}, \end{aligned} \tag{3}$$

where, $\overline{\mathbb{C}}$ is the effective stiffness tensor, $\overline{\mathbb{S}}$ is the effective compliance tensor, $\boldsymbol{E}_\theta$ is the effective thermal strain tensor and $\boldsymbol{B}\big(= \overline{\mathbb{C}}[\boldsymbol{E}_\theta]\big)$. The terms $\boldsymbol{E}$ is the effective strain or macroscopic strain and $\boldsymbol{\Sigma}$ is the effective stress or macroscopic stress. At the macroscale, either $\boldsymbol{E}$ or $\boldsymbol{\Sigma}$ is usually prescribed at a point. This serves as the boundary condition for the RVE of the considered microstructure geometry of the random heterogeneous material.

The effective stiffness in Equation (3) is evaluated using the Walpole notation for singular approximation [16–19] which is given as,

$$\overline{\mathbb{C}} = \langle \mathbb{L} \rangle^{-1} + \mathbb{C}_\mathrm{o} - \mathbb{P}_\mathrm{o}^{-1}, \qquad \mathbb{L} = (\mathbb{C} - \mathbb{C}_\mathrm{o} + \mathbb{P}_\mathrm{o}^{-1})^{-1}. \tag{4}$$

where $\mathbb{C}_\text{o}$ is the reference stiffness. Different choices of $\mathbb{C}_\text{o}$ leads to different estimates. The Hill's polarization tensor $\mathbb{P}_\text{o}$ is expressed as integral over the surface of an ellipsoid which is given as,

$$\mathbb{P}_\text{o} = \frac{1}{4\pi}\int_S \mathbb{H}(\mathbb{C}_\text{o}, \boldsymbol{n}) \left(\frac{\boldsymbol{n}\cdot(\boldsymbol{A}^{-\text{T}}\boldsymbol{A}^{-1}\boldsymbol{n})}{\det(\boldsymbol{A})^{-\frac{2}{3}}}\right)^{-\frac{3}{2}} \mathrm{d}S,$$
$$\mathbb{H}(\mathbb{C}_\text{o}, \boldsymbol{n}) = \mathbb{I}^\text{s}(\boldsymbol{N}\times\boldsymbol{K}^{-1})\mathbb{I}^\text{s}, \qquad \boldsymbol{K} = \big(\mathbb{I}\times(\boldsymbol{n}\otimes\boldsymbol{n})\big)[\mathbb{C}_\text{o}], \qquad \boldsymbol{A} = \sqrt{\boldsymbol{Z}}. \tag{5}$$

where, $\boldsymbol{n}$ is the unit normal vector given in a spherical coordinate system i.e., in matrix representation $[\boldsymbol{n}] = [\sin\theta\cos\varphi\,,\quad \sin\theta\sin\varphi\,,\quad \cos\theta]^\text{T}$ (see Fig. 2). The second order tensor, $\boldsymbol{N} = \boldsymbol{n}\otimes\boldsymbol{n}$ and $\boldsymbol{Z}$ is the ellipsoidal shape of the inhomogeneity. The matrix representation of $\boldsymbol{Z}$ is,

$$[\boldsymbol{Z}] = \begin{bmatrix} 1/d & & \\ & 1/d & \\ & & 1/ad \end{bmatrix}.$$

The fourth order tensor $\mathbb{H}(\mathbb{C}_\text{o}, \boldsymbol{n})$ with major and minor symmetric is expressed in terms of acoustic or Christoffel tensor $\boldsymbol{K}$. The fourth order symmetric tensor $\mathbb{H}(\mathbb{C}_\text{o}, \boldsymbol{n})$ is expressed in terms of acoustic or Christoffel tensor $\boldsymbol{K}$.

Considering the advantage of the linearity of the phase constitutive relations local stress and total strain fields in a linear thermoelastic homogenization problem can be decomposed for calculation of $\boldsymbol{E}_\theta$ in Eq. (3). Fig. 1 schematically illustrates the decomposition of the thermoelastic problem into two subproblems. Problem 1 is a purely elastic analysis (field quantities indicated by $(\bullet)^\text{e}$) subjected to a homogeneous traction boundary condition. Conversely, one can solve this linear problem by specifying the displacement on the boundary using periodic boundary condition for e.g., in a finite element framework. In any case, they satisfy the Hill-Mandel condition. Problem 2 is a homogeneous, stress-free strain boundary value problem i.e., a free expansion problem due to temperature change only (field quantities indicated by $(\bullet)^\text{t}$).

The averaged field quantities and constraints of the problems is summarized in the Table 1 [20]. For a 2-phase composite, $\boldsymbol{E}_\theta$ can be computed using the outcomes of the solution of problem 1 and 2 [21], [22] which is given as,

$$\boldsymbol{E}_\theta = \langle \boldsymbol{\varepsilon}_\theta \rangle + \big(\overline{\mathbb{S}} - \langle \mathbb{S} \rangle\big)(\Delta\mathbb{S})^{-1}[\Delta\boldsymbol{\varepsilon}_\theta], \tag{6}$$

where, $\Delta = (\cdot)_2 - (\cdot)_1$ and $\langle \bullet \rangle = \sum_{\gamma=1}^{2} c_\gamma (\bullet)_\gamma$

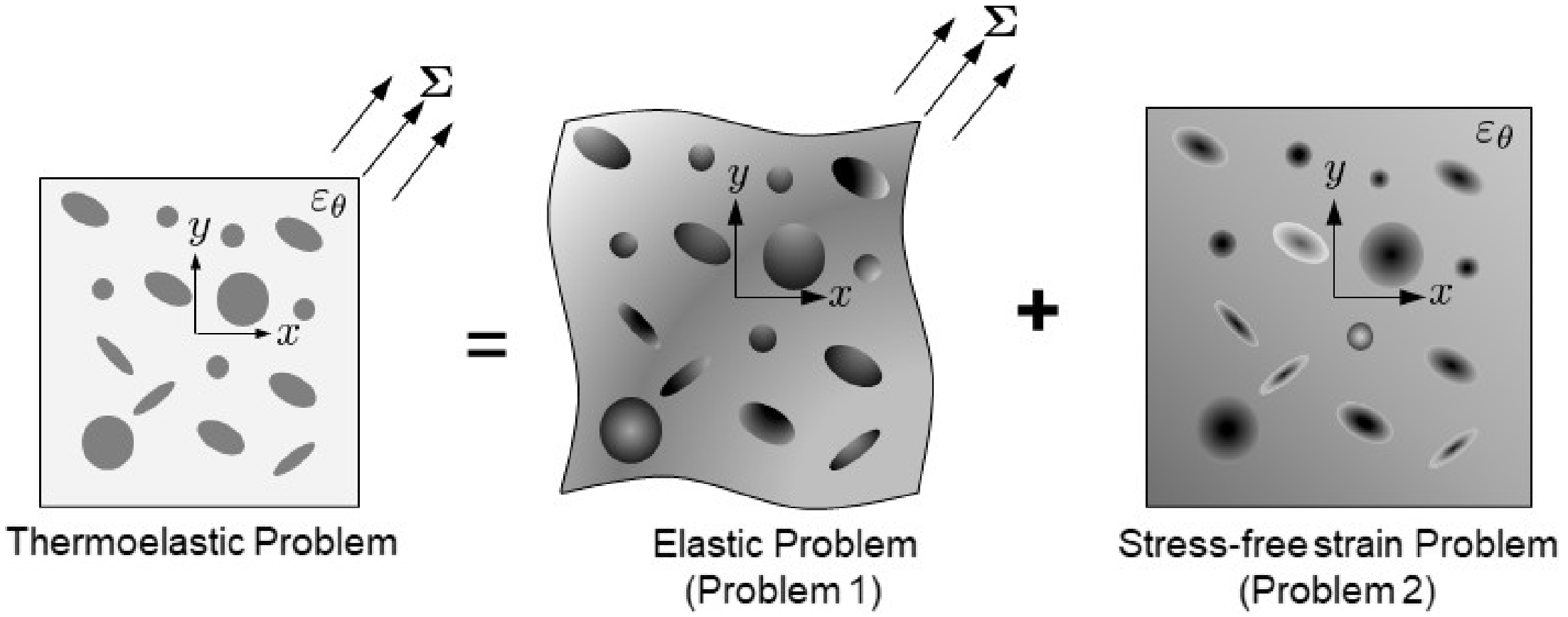

**Fig. 1.** Schematic of decomposition of the thermoelastic problem. The gradient of grey indicates stress fluctuation field induced due to mechanical load and stress-free strain loading.

## 3 Statistical Moments of the Local Fields

### 3.1 First Statistical Moments (Mean)

Due to the uniqueness of the solution for linear elastic boundary value problems, a unique and exact fourth-order localization (or concentration) tensor $\mathbb{A}_\gamma(\boldsymbol{x})$ and $\mathbb{B}_\gamma(\boldsymbol{x})$ can be defined, linking the microscale and macroscale strain and stress quantities respectively. In a linear thermoelastic problem, due to the mismatch in thermal expansion coefficients of the phases, an additional second-order thermal strain and stress concentration tensor, $\boldsymbol{a}_\gamma(\boldsymbol{x})$ and $\boldsymbol{b}_\gamma(\boldsymbol{x})$ appears [20]. The averaged concentration tensors with following restrictions yield the first moments from the macroscopic scale quantities,

$$\begin{aligned} \langle \boldsymbol{\varepsilon}(\boldsymbol{x}) \rangle_\gamma &= \mathbb{A}_\gamma[\boldsymbol{E}] + \boldsymbol{a}_\gamma, \qquad \langle \mathbb{A} \rangle = \mathbb{I}^{\mathrm{s}} \text{ and } \langle \boldsymbol{a} \rangle = \boldsymbol{0}, \\ \langle \boldsymbol{\sigma}(\boldsymbol{x}) \rangle_\gamma &= \mathbb{B}_\gamma[\boldsymbol{\Sigma}] + \boldsymbol{b}_\gamma, \qquad \langle \mathbb{B} \rangle = \mathbb{I}^{\mathrm{s}} \text{ and } \langle \boldsymbol{b} \rangle = \boldsymbol{0}. \end{aligned} \tag{7}$$

**Table 1.** Resulting quantities and constraints of the thermoelastic problems.

| Field Quantity | Elastic Problem | Stress-Free Strain Problem | Linear Thermoelastic Problem |
|---|---|---|---|
| $\boldsymbol{\sigma}(\boldsymbol{x})$ and $\boldsymbol{\varepsilon}(\boldsymbol{x})$ | $\boldsymbol{\sigma}^{\mathrm{e}}(\boldsymbol{x}) = \mathbb{C}[\boldsymbol{\varepsilon}^{\mathrm{e}}(\boldsymbol{x})]$<br>$\boldsymbol{\varepsilon}^{\mathrm{e}}(\boldsymbol{x}) = \mathbb{S}[\boldsymbol{\sigma}^{\mathrm{e}}(\boldsymbol{x})]$ | $\boldsymbol{\sigma}^{\mathrm{t}}(\boldsymbol{x}) = \mathbb{C}[\boldsymbol{\varepsilon}^{\mathrm{t}}(\boldsymbol{x})] + \boldsymbol{\beta}(\boldsymbol{x})$<br>$\boldsymbol{\varepsilon}^{\mathrm{t}}(\boldsymbol{x}) = \mathbb{S}[\boldsymbol{\sigma}^{\mathrm{t}}(\boldsymbol{x})] + \boldsymbol{\varepsilon}_\theta$ | $\boldsymbol{\sigma}(\boldsymbol{x}) = \boldsymbol{\sigma}^{\mathrm{e}}(\boldsymbol{x}) + \boldsymbol{\sigma}^{\mathrm{t}}(\boldsymbol{x})$<br>$\boldsymbol{\varepsilon}(\boldsymbol{x}) = \boldsymbol{\varepsilon}^{\mathrm{e}}(\boldsymbol{x}) + \boldsymbol{\varepsilon}^{\mathrm{t}}(\boldsymbol{x})$ |
| $\boldsymbol{\beta}(\boldsymbol{x})$ and $\boldsymbol{\varepsilon}_\theta$ | $\boldsymbol{\beta}(\boldsymbol{x}) = \boldsymbol{\varepsilon}_\theta = \boldsymbol{0}$ | $\boldsymbol{\beta}(\boldsymbol{x}) \neq \boldsymbol{\varepsilon}_\theta \neq \boldsymbol{0}$ | $\langle \boldsymbol{\varepsilon}_\theta \rangle \neq \boldsymbol{E}_\theta$ |
| $\boldsymbol{\Sigma}$ | $\langle \boldsymbol{\sigma}^{\mathrm{e}} \rangle = \boldsymbol{\Sigma}$ | $\langle \boldsymbol{\sigma}^{\mathrm{t}} \rangle = \boldsymbol{0}$ | $\boldsymbol{\Sigma} = \bar{\mathbb{C}}[\boldsymbol{E}] - \boldsymbol{B}$ |
| $\boldsymbol{E}$ | $\boldsymbol{E} = \langle \boldsymbol{\varepsilon}^{\mathrm{e}} \rangle = \bar{\mathbb{S}}[\boldsymbol{\Sigma}]$ | $\langle \boldsymbol{\varepsilon}^{\mathrm{t}} \rangle = \boldsymbol{E}_\theta$ | $\boldsymbol{E} = \bar{\mathbb{S}}[\boldsymbol{\Sigma}] + \boldsymbol{E}_\theta$ |
| $W$ | $W^{\mathrm{e}} = \frac{1}{2}\boldsymbol{\Sigma} \cdot \boldsymbol{E}$<br>$= \frac{1}{2}\langle \boldsymbol{\sigma}^{\mathrm{e}} \cdot \boldsymbol{\varepsilon}^{\mathrm{e}} \rangle$ | $W^{\mathrm{t}} = \frac{1}{2}\langle \boldsymbol{\sigma}^{\mathrm{t}} \cdot (\boldsymbol{\varepsilon}^{\mathrm{t}} - \boldsymbol{\varepsilon}_\theta) \rangle$<br>$= -\frac{1}{2}\langle \boldsymbol{\sigma}^{\mathrm{t}} \cdot \boldsymbol{\varepsilon}_\theta \rangle$ | $W = W^{\mathrm{e}} + W^{\mathrm{t}}$<br>$= \frac{1}{2}\langle \boldsymbol{\sigma} \cdot (\boldsymbol{\varepsilon} - \boldsymbol{\varepsilon}_\theta) \rangle$<br>$= \frac{1}{2}\langle \boldsymbol{\sigma}^{\mathrm{e}} \cdot \boldsymbol{\varepsilon}^{\mathrm{e}} \rangle - \frac{1}{2}\langle \boldsymbol{\sigma}^{\mathrm{t}} \cdot \boldsymbol{\varepsilon}_\theta \rangle$ |

The relation among the concentration tensors is given as,

$$\mathbb{A}_\gamma = \mathbb{S}_\gamma \mathbb{B}_\gamma \overline{\mathbb{C}}, \quad \mathbb{B}_\gamma = \mathbb{C}_\gamma \mathbb{A}_\gamma \overline{\mathbb{S}}$$

The effective elastic properties of the composite [23] using localization tensors is then defined as,

$$\overline{\mathbb{C}} = \sum_{\gamma=1}^{n} c_\gamma \mathbb{C}_\gamma \mathbb{A}_\gamma, \quad \overline{\mathbb{S}} = \sum_{\gamma=1}^{n} c_\gamma \mathbb{S}_\gamma \mathbb{B}_\gamma. \tag{8}$$

In the linear elastic problem, if the strain concentration tensor is defined then the effective stiffness and the first moments of local fields can be computed e.g., in the case of Mori-Tanaka method. In other words, the averaged localization tensor and the first moments of local fields can be computed if the effective stiffness is known e.g., using full-field homogenization methods. For a two-phase composite using Eq. (8) and the identities in Eq. (7), the localization tensors in terms of effective stiffness [20] can be expressed as,

$$\begin{aligned} \mathbb{A}_\gamma &= \mathbb{I}^{\mathrm{s}} - \frac{1}{c_\gamma}(\Delta\mathbb{C})^{-\mathrm{T}_H}\left(\overline{\mathbb{C}} - \langle\mathbb{C}\rangle\right)^{\mathrm{T}_H}, \\ \boldsymbol{a}_\gamma &= \left(\mathbb{A}_\gamma - \mathbb{I}^{\mathrm{s}}\right)\frac{1}{c_\gamma}(\Delta\mathbb{C})^{-1}(\Delta\boldsymbol{\beta})^{\mathrm{T}}. \end{aligned} \tag{9}$$

where $\Delta\boldsymbol{\beta} = \mathbb{C}_2\boldsymbol{\varepsilon}_{\theta 2} - \mathbb{C}_1\boldsymbol{\varepsilon}_{\theta 1}$. Within the framework of mean-field homogenization, the solution of concentration tensor for one elastic phase is sufficient to characterize the full thermoelastic behavior of a two-phase heterogenous material (see Eq. (9) and (7).

### 3.2 Second Statistical Moments

The second moment of the stress field $(\boldsymbol{\sigma} = \boldsymbol{\sigma}^{\mathrm{e}} + \boldsymbol{\sigma}^{\mathrm{t}})$ for $\gamma-$ phase of composite can be evaluated by summing the second moments of stress field generated due to the decomposed problem and the interaction of the stress fields of both the problems. The expression is given as,

$$\langle \boldsymbol{\sigma}^{\otimes 2} \rangle_\gamma = \langle (\boldsymbol{\sigma}^{\mathrm{e}})^{\otimes 2} \rangle_\gamma + \langle (\boldsymbol{\sigma}^{\mathrm{t}})^{\otimes 2} \rangle_\gamma + \langle \boldsymbol{\sigma}^{\mathrm{e}} \otimes \boldsymbol{\sigma}^{\mathrm{t}} \rangle_\gamma + \langle \boldsymbol{\sigma}^{\mathrm{t}} \otimes \boldsymbol{\sigma}^{\mathrm{e}} \rangle_\gamma. \tag{10}$$

### 3.2.1 Derivation of the term $\langle (\boldsymbol{\sigma}^{\mathrm{e}})^{\otimes 2} \rangle_\gamma$.

Consider the effective energy term $W^{\mathrm{e}}$ due to the linear elastic problem 1 as given in the Table 1,

$$W^{\mathrm{e}} = \frac{1}{2} \langle \boldsymbol{\sigma}^{\mathrm{e}} \cdot \boldsymbol{\varepsilon}^{\mathrm{e}} \rangle = \frac{1}{2} \langle \mathbb{S} \cdot (\boldsymbol{\sigma}^{\mathrm{e}})^{\otimes 2} \rangle. \tag{11}$$

A perturbation on both left- and right-hand side of the equation would result in,

$$\begin{aligned} \delta W^{\mathrm{e}} &= \frac{1}{2} (\langle \delta\boldsymbol{\sigma}^{\mathrm{e}} \cdot \mathbb{S}[\boldsymbol{\sigma}^{\mathrm{e}}] \rangle + \langle \boldsymbol{\sigma}^{\mathrm{e}} \cdot \delta\mathbb{S}[\boldsymbol{\sigma}^{\mathrm{e}}] \rangle + \langle \boldsymbol{\sigma}^{\mathrm{e}} \cdot \mathbb{S}[\delta\boldsymbol{\sigma}^{\mathrm{e}}] \rangle) \\ &= \frac{1}{2} (2\langle \delta\boldsymbol{\sigma}^{\mathrm{e}} \cdot \mathbb{S}[\boldsymbol{\sigma}^{\mathrm{e}}] \rangle + \langle \boldsymbol{\sigma}^{\mathrm{e}} \cdot \delta\mathbb{S}[\boldsymbol{\sigma}^{\mathrm{e}}] \rangle). \end{aligned} \tag{12}$$

Since the both old and new stress fields have to satisfy the same homogenous traction boundary condition, hence $\langle \delta\boldsymbol{\sigma}^{\mathrm{e}} \rangle = \mathbf{0}$. Hence, the conditions $\langle \delta\boldsymbol{\sigma}^{\mathrm{e}} \cdot \mathbb{S}_\gamma[\boldsymbol{\sigma}^{\mathrm{e}}] \rangle = 0$ must hold true to satisfy the Hill-Mandel condition. Equation (12) can be sorted as,

$$\delta W^{\mathrm{e}} = \frac{1}{2} \langle \boldsymbol{\sigma}^{\mathrm{e}} \cdot \delta\mathbb{S}[\boldsymbol{\sigma}^{\mathrm{e}}] \rangle.$$

If the compliance tensor of $\gamma$-phase is varied by $\delta\mathbb{S}_\gamma$ keeping the compliance tensor of other phases and the resulting effective compliance same [7, 19], then the variation in the effective energy would yield as,

$$\delta W^{\mathrm{e}} = \frac{c_\gamma}{2} \langle (\boldsymbol{\sigma}^{\mathrm{e}})^{\otimes 2} \rangle_\gamma \cdot \delta\mathbb{S}_\gamma. \tag{13}$$

Using variational derivative and using Eq. (13),

$$\delta W^{\mathrm{e}} = \frac{\partial W^{\mathrm{e}}}{\partial \mathbb{S}_\gamma} \cdot \delta\mathbb{S}_\gamma = \frac{c_\gamma}{2} \langle (\boldsymbol{\sigma}^{\mathrm{e}})^{\otimes 2} \rangle_\gamma \cdot \delta\mathbb{S}_\gamma,$$

The following relation will hold true for any arbitrary choice of $\delta\mathbb{S}_\gamma$,

$$\langle(\boldsymbol{\sigma}^{\mathrm{e}})^{\otimes 2}\rangle_\gamma = \frac{2}{c_\gamma}\frac{\partial W^{\mathrm{e}}}{\partial \mathbb{S}_\gamma}. \tag{14}$$

The computation of $\langle(\boldsymbol{\sigma}^{\mathrm{e}})^{\otimes 2}\rangle$ in Eq. (14) can be evaluated by taking the anisotropic energy derivatives with phase stiffness $\mathbb{S}_\gamma$. In general, for any material symmetry of $\mathbb{S}_\gamma$ taking $\partial W^{\mathrm{e}}/\partial \mathbb{S}_\gamma$ (see Table 1) yields,

$$\langle(\boldsymbol{\sigma}^{\mathrm{e}})^{\otimes 2}\rangle_\gamma = \frac{2}{c_\gamma}\frac{\partial W^{\mathrm{e}}}{\partial \mathbb{S}_\gamma} = \frac{1}{c_\gamma}\left(\frac{\partial \bar{\mathbb{S}}}{\partial \mathbb{S}_\gamma}\right)^{\mathrm{T_H}}[(\boldsymbol{\Sigma})^{\otimes 2}] = \mathbb{C}_\gamma\langle(\boldsymbol{\varepsilon}^{\mathrm{e}})^{\otimes 2}\rangle_\gamma\mathbb{C}_\gamma. \tag{15}$$

The term $\partial\bar{\mathbb{S}}/\partial\mathbb{S}_\gamma$ can be evaluated from the effective stiffness which is given as,

$$\frac{\partial \bar{\mathbb{S}}}{\partial \mathbb{S}_\gamma} = \bar{\mathbb{S}}^{\times 2}\frac{\partial \bar{\mathbb{C}}}{\partial \mathbb{C}_\gamma}[\mathbb{C}_\gamma^{\times 2}\boldsymbol{\mathcal{J}}^{\mathrm{s}}]. \tag{16}$$

In index notation Eq. (16) can be written as,

$$\frac{\partial \bar{S}_{ijkl}}{\partial S^\gamma_{mnop}} = \bar{S}_{ijab}\left(\frac{\partial \bar{C}}{\partial C_\gamma}\right)_{abcdefgh}\bar{S}_{cdkl}C^\gamma_{efuv}\delta^{8\mathrm{s}}_{uvwxmnop}C^\gamma_{wxgh}. \tag{17}$$

The derivatives of the effective stiffness with phase stiffness for any type of anisotropy can be determined using Eq. (4). It is given as,

$$\begin{aligned}\frac{\partial \bar{\mathbb{C}}}{\partial \mathbb{C}_\gamma} &= \frac{\partial\langle\mathbb{L}\rangle^{-1}}{\partial \mathbb{C}_\gamma} + \frac{\partial \mathbb{C}_{\mathrm{o}}}{\partial \mathbb{C}_\gamma} - \frac{\partial \mathbb{P}_{\mathrm{o}}^{-1}}{\partial \mathbb{C}_{\mathrm{o}}}\frac{\partial \mathbb{C}_{\mathrm{o}}}{\partial \mathbb{C}_\gamma},\\ \Rightarrow \frac{\partial \bar{\mathbb{C}}}{\partial \mathbb{C}_\gamma} &= -(\langle\mathbb{L}\rangle^{-1})^{\times 2}\frac{\partial\langle\mathbb{L}\rangle}{\partial \mathbb{C}_\gamma} + \frac{\partial \mathbb{C}_{\mathrm{o}}}{\partial \mathbb{C}_\gamma} - (\mathbb{P}_{\mathrm{o}}^{-1})^{\times 2}\frac{\partial \mathbb{P}_{\mathrm{o}}}{\partial \mathbb{C}_{\mathrm{o}}}\frac{\partial \mathbb{C}_{\mathrm{o}}}{\partial \mathbb{C}_\gamma}.\end{aligned} \tag{18}$$

Derivative of the terms $\partial\langle\mathbb{L}\rangle/\partial\mathbb{C}_\gamma$ involved in Eq. (18) can be expressed as,

$$\frac{\partial\langle\mathbb{L}\rangle}{\partial \mathbb{C}_\gamma} = -\sum_{\lambda=1}^{2} c_\lambda \mathbb{L}_\lambda^{\times 2}\left(\delta_{\lambda\gamma}\boldsymbol{\mathcal{J}}^{\mathrm{s}} - \frac{\partial \mathbb{C}_{\mathrm{o}}}{\partial \mathbb{C}_\gamma} - (\mathbb{P}_{\mathrm{o}}^{-1})^{\times 2}\frac{\partial \mathbb{P}_{\mathrm{o}}}{\partial \mathbb{C}_{\mathrm{o}}}\frac{\partial \mathbb{C}_{\mathrm{o}}}{\partial \mathbb{C}_\gamma}\right), \tag{19}$$

where $\mathbb{L}_\lambda = (\mathbb{C}_\lambda - \mathbb{C}_o + \mathbb{P}_o^{-1})^{-1}$. In Eq. (19) if $\partial \mathbb{P}_o / \partial \mathbb{C}_o$ is known, then $\partial \overline{\mathbb{S}} \big/ \partial \mathbb{S}_\gamma$ can be found via Eq. (16) and (18) to find Eq. (15). These derivatives are evaluated considering the stiffness tensor is completely anisotropic with major and minor symmetric property.

### 3.2.2 Derivation of the term $\langle (\boldsymbol{\sigma}^{\mathbf{t}})^{\otimes 2} \rangle_\gamma$

Consider the energy term due to the linear elastic problem 2 as given in the Table 1 and using Hill-Mandel condition,

$$W^{\mathrm{t}} = \frac{1}{2} \langle \boldsymbol{\sigma}^{\mathrm{t}} \cdot (\boldsymbol{\varepsilon}^{\mathrm{t}} - \boldsymbol{\varepsilon}_\theta) \rangle = \frac{1}{2} \langle \mathbb{S} \cdot (\boldsymbol{\sigma}^{\mathrm{t}})^{\otimes 2} \rangle = -\frac{1}{2} \langle \boldsymbol{\sigma}^{\mathrm{t}} \cdot \boldsymbol{\varepsilon}_\theta \rangle. \tag{20}$$

A perturbation on both left- and right-hand side of the equation $(20)_2$ considering $\boldsymbol{\varepsilon}_\theta$ is homogenous in each phase would result in,

$$2\delta W^{\mathrm{t}} = (2\langle \delta\boldsymbol{\sigma}^{\mathrm{t}} \cdot \mathbb{S}[\boldsymbol{\sigma}^{\mathrm{t}}] \rangle + \langle \boldsymbol{\sigma}^{\mathrm{t}} \cdot \delta\mathbb{S}[\boldsymbol{\sigma}^{\mathrm{t}}] \rangle) = -\langle \delta\boldsymbol{\sigma}^{\mathrm{t}} \cdot \boldsymbol{\varepsilon}_\theta \rangle. \tag{21}$$

Using the constitutive relation of the phase Eq. $(1)_1$ in (21),

$$2\delta W^{\mathrm{t}} = (2\langle \delta\boldsymbol{\sigma}^{\mathrm{t}} \cdot \boldsymbol{\varepsilon}^{\mathrm{t}} \rangle - 2\langle \delta\boldsymbol{\sigma}^{\mathrm{t}} \cdot \boldsymbol{\varepsilon}_\theta \rangle + \langle \delta\mathbb{S} \cdot (\boldsymbol{\sigma}^{\mathrm{t}})^{\otimes 2} \rangle) = -\langle \delta\boldsymbol{\sigma}^{\mathrm{t}} \cdot \boldsymbol{\varepsilon}_\theta \rangle. \tag{22}$$

The both old and new stress fields have to satisfy the same zero traction boundary condition, hence $\langle \delta\boldsymbol{\sigma}^{\mathrm{t}} \rangle = \langle \boldsymbol{\sigma}^{\mathrm{t}} \rangle = \mathbf{0}$ and $\langle \boldsymbol{\sigma}^{\mathrm{t}} \cdot \boldsymbol{\varepsilon}^{\mathrm{t}} \rangle = \langle \delta\boldsymbol{\sigma}^{\mathrm{t}} \cdot \boldsymbol{\varepsilon}^{\mathrm{t}} \rangle = 0$ following the Hill-Mandel condition. Eq. (22) can be re-written as,

$$2\delta W^{\mathrm{t}} = -2\langle \delta\boldsymbol{\sigma}^{\mathrm{t}} \cdot \boldsymbol{\varepsilon}_\theta \rangle + \langle \delta\mathbb{S} \cdot (\boldsymbol{\sigma}^{\mathrm{t}})^{\otimes 2} \rangle. \tag{23}$$

Using Eq. $(22)_3$ in Eq. (23),

$$\langle \delta\mathbb{S} \cdot (\boldsymbol{\sigma}^{\mathrm{t}})^{\otimes 2} \rangle = -2\delta W^{\mathrm{t}}. \tag{24}$$

If the compliance tensor of $\gamma$-phase is varied by $\delta\mathbb{S}_\gamma$ keeping the compliance tensor of other phases and the resulting effective compliance fixed, then $\delta W^{\mathrm{t}}$ would yield as

$$\delta W^{\mathrm{t}} = -\frac{c_{\gamma}}{2}\langle(\boldsymbol{\sigma}^{\mathrm{t}})^{\otimes 2}\rangle_{\gamma}\cdot\delta\mathbb{S}_{\gamma}. \tag{25}$$

Using the variational derivative and using Equation (25)

$$\delta W^{\mathrm{t}} = \frac{\partial W^{\mathrm{t}}}{\partial\mathbb{S}_{\gamma}}\cdot\delta\mathbb{S}_{\gamma} = -\frac{c_{\gamma}}{2}\langle(\boldsymbol{\sigma}^{\mathrm{t}})^{\otimes 2}\rangle_{\gamma}\cdot\delta\mathbb{S}_{\gamma}. \tag{26}$$

The following relation will hold true for any arbitrary choice of $\delta\mathbb{S}_{\gamma}$

$$\langle(\boldsymbol{\sigma}^{\mathrm{t}})^{\otimes 2}\rangle_{\gamma} = -\frac{2}{c_{\gamma}}\frac{\partial W^{\mathrm{t}}}{\partial\mathbb{S}_{\gamma}}. \tag{27}$$

For a two-phase composite the expression for $W^{\mathrm{t}}$, using Eq. $(20)_3$ can be expressed in terms of $\overline{\mathbb{S}}$ or $\boldsymbol{E}_{\theta}$ [22] as,

$$\begin{aligned} W^{\mathrm{t}} = -\frac{1}{2}\langle\boldsymbol{\sigma}^{\mathrm{t}}\cdot\boldsymbol{\varepsilon}_{\theta}\rangle &= \frac{1}{2}\Delta\boldsymbol{\varepsilon}_{\theta}(\Delta\mathbb{S})^{-1}\big(\langle\mathbb{S}\rangle-\overline{\mathbb{S}}\big)(\Delta\mathbb{S})^{-1}\Delta\boldsymbol{\varepsilon}_{\theta} \\ &= \frac{1}{2}\Delta\boldsymbol{\varepsilon}_{\theta}\cdot((\Delta\mathbb{S})^{-1}[\langle\boldsymbol{\varepsilon}_{\theta}\rangle-\boldsymbol{E}_{\theta}]). \end{aligned} \tag{28}$$

Using Eq. $(28)_3$,

$$\frac{\partial W^{\mathrm{t}}}{\partial\mathbb{S}_{\gamma}} = \frac{1}{2}\left((((\Delta\mathbb{S})^{-1})^{\times 2}\boldsymbol{\mathcal{J}}^{\mathrm{s}})^{\mathrm{T_H}}[\Delta\boldsymbol{\varepsilon}_{\theta}\otimes(\langle\boldsymbol{\varepsilon}_{\theta}\rangle-\boldsymbol{E}_{\theta})]-(\Delta\mathbb{S}^{-1}[\Delta\boldsymbol{\varepsilon}_{\theta}])\frac{\partial\boldsymbol{E}_{\theta}}{\partial\mathbb{S}_{\gamma}}\right). \tag{29}$$

In index notation, Eq. (29) can be re-written as,

$$\left(\frac{\partial W^{\mathrm{t}}}{\partial\mathbb{S}_{\gamma}}\right)_{ijkl} = \frac{1}{2}\Delta\varepsilon^{\theta}_{ab}\left(\Delta S^{-1}_{abpq}\delta^{8\mathrm{s}}_{pqrsijkl}\Delta S^{-1}_{rscd}(\langle\boldsymbol{\varepsilon}_{\theta}\rangle_{cd}-E^{\theta}_{cd})-\Delta S^{-1}_{abcd}\frac{\partial E^{\theta}_{cd}}{\partial S^{\gamma}_{ijkl}}\right). \tag{30}$$

The unknown term in the equation is $\partial\boldsymbol{E}_{\theta}/\partial\mathbb{S}_{\gamma}$ which can be computed using anisotropic derivatives of Eq. (6).

### 3.2.3 Derivation of term $\langle \boldsymbol{\sigma}^{\mathbf{e}} \otimes \boldsymbol{\sigma}^{\mathbf{t}} \rangle_{\gamma}$.

***Interaction of $\boldsymbol{\sigma}^{\mathbf{t}}$ and $\boldsymbol{\varepsilon}^{\mathrm{e}}$ fields***

Applying variation to interaction of the field quantities $\boldsymbol{\sigma}^{\mathrm{t}}$ and $\boldsymbol{\varepsilon}^{\mathrm{e}}$,

$$\langle \boldsymbol{\sigma}^{\mathrm{t}} \cdot \boldsymbol{\varepsilon}^{\mathrm{e}} \rangle = \langle \boldsymbol{\sigma}^{\mathrm{t}} \cdot \mathbb{S}[\boldsymbol{\sigma}^{\mathrm{e}}] \rangle = \mathbf{0}, \tag{31}$$

the expression can be written as,

$$\begin{aligned} \langle \delta\boldsymbol{\sigma}^{\mathrm{t}} \cdot \mathbb{S}[\boldsymbol{\sigma}^{\mathrm{e}}] \rangle + \langle \boldsymbol{\sigma}^{\mathrm{t}} \cdot \delta\mathbb{S}[\boldsymbol{\sigma}^{\mathrm{e}}] \rangle + \langle \boldsymbol{\sigma}^{\mathrm{t}} \cdot \mathbb{S}[\delta\boldsymbol{\sigma}^{\mathrm{e}}] \rangle = 0 \\ = \langle \delta\boldsymbol{\sigma}^{\mathrm{t}} \cdot \mathbb{S}[\boldsymbol{\sigma}^{\mathrm{e}}] \rangle + \langle \boldsymbol{\sigma}^{\mathrm{t}} \cdot \delta\mathbb{S}[\boldsymbol{\sigma}^{\mathrm{e}}] \rangle + \langle \delta\boldsymbol{\sigma}^{\mathrm{e}} \cdot \mathbb{S}[\boldsymbol{\sigma}^{\mathrm{t}}] \rangle = \\ = \langle \delta\boldsymbol{\sigma}^{\mathrm{t}} \cdot \mathbb{S}[\boldsymbol{\sigma}^{\mathrm{e}}] \rangle + \langle \boldsymbol{\sigma}^{\mathrm{t}} \cdot \delta\mathbb{S}[\boldsymbol{\sigma}^{\mathrm{e}}] \rangle + \langle \delta\boldsymbol{\sigma}^{\mathrm{e}} \cdot (\boldsymbol{\varepsilon}^{\mathrm{t}} - \boldsymbol{\varepsilon}_{\theta}) \rangle = \mathbf{0}. \end{aligned} \tag{32}$$

The conditions of the problem 1 and problem 2 are $\langle \delta\boldsymbol{\sigma}^{\mathrm{e}} \rangle = \mathbf{0}$ and $\langle \delta\boldsymbol{\sigma}^{\mathrm{t}} \rangle = \langle \boldsymbol{\sigma}^{\mathrm{t}} \rangle = \mathbf{0}$ respectively. Using these conditions and Hill-Mandel condition,

$$\begin{aligned} \langle \delta\boldsymbol{\sigma}^{\mathrm{t}} \cdot \mathbb{S}[\boldsymbol{\sigma}^{\mathrm{e}}] \rangle = \langle \delta\boldsymbol{\sigma}^{\mathrm{t}} \rangle \cdot \langle \mathbb{S}[\boldsymbol{\sigma}^{\mathrm{e}}] \rangle = \mathbf{0}, \\ \langle \delta\boldsymbol{\sigma}^{\mathrm{e}} \cdot \boldsymbol{\varepsilon}^{\mathrm{t}} \rangle = \langle \delta\boldsymbol{\sigma}^{\mathrm{e}} \rangle \cdot \langle \boldsymbol{\varepsilon}^{\mathrm{t}} \rangle = 0. \end{aligned} \tag{33}$$

Hence Eq. (32) can be written as,

$$\langle \boldsymbol{\sigma}^{\mathrm{t}} \cdot \delta\mathbb{S}[\boldsymbol{\sigma}^{\mathrm{e}}] \rangle - \langle \delta\boldsymbol{\sigma}^{\mathrm{e}} \cdot \boldsymbol{\varepsilon}_{\theta} \rangle = 0. \tag{34}$$

***Interaction of $\boldsymbol{\sigma}^{\mathbf{e}}$ and $\boldsymbol{\varepsilon}^{\mathbf{t}}$ fields***

Consider the interaction of the field quantities $\boldsymbol{\sigma}^{\mathrm{e}}$ and $\boldsymbol{\varepsilon}^{\mathrm{t}}$, and using Hill-Mandel condition on these quantities,

$$\langle \boldsymbol{\sigma}^{\mathrm{e}} \cdot \boldsymbol{\varepsilon}^{\mathrm{t}} \rangle = \langle \boldsymbol{\sigma}^{\mathrm{e}} \rangle \cdot \langle \boldsymbol{\varepsilon}^{\mathrm{t}} \rangle = \boldsymbol{\Sigma} \cdot \boldsymbol{E}_{\theta}. \tag{35}$$

Applying variation to the left side of the equation yields to,

$$\langle \delta\boldsymbol{\sigma}^{\mathrm{e}} \cdot \boldsymbol{\varepsilon}^{\mathrm{t}} \rangle + \langle \boldsymbol{\sigma}^{\mathrm{e}} \cdot \delta\boldsymbol{\varepsilon}^{\mathrm{t}} \rangle = \langle \boldsymbol{\sigma}^{\mathrm{e}} \rangle \cdot \langle \delta\boldsymbol{\varepsilon}^{\mathrm{t}} \rangle = \boldsymbol{\Sigma} \cdot \delta\boldsymbol{E}_{\theta}. \tag{36}$$

Adding and subtracting $\boldsymbol{\varepsilon}_{\theta}$ in Eq. (35) and using Hill-Mandel condition, it can be rewritten as,

$$\begin{aligned} \langle \boldsymbol{\sigma}^{\mathrm{e}} \cdot \boldsymbol{\varepsilon}^{\mathrm{t}} \rangle &= \langle \boldsymbol{\sigma}^{\mathrm{e}} \cdot (\boldsymbol{\varepsilon}^{\mathrm{t}} - \boldsymbol{\varepsilon}_{\theta} + \boldsymbol{\varepsilon}_{\theta}) \rangle = \langle \boldsymbol{\sigma}^{\mathrm{e}} \cdot (\boldsymbol{\varepsilon}^{\mathrm{t}} - \boldsymbol{\varepsilon}_{\theta}) \rangle + \langle \boldsymbol{\sigma}^{\mathrm{e}} \cdot \boldsymbol{\varepsilon}_{\theta} \rangle \\ &= \langle \boldsymbol{\varepsilon}^{\mathrm{e}} \cdot \boldsymbol{\sigma}^{\mathrm{t}} \rangle + \langle \boldsymbol{\sigma}^{\mathrm{e}} \cdot \boldsymbol{\varepsilon}_{\theta} \rangle = \langle \boldsymbol{\varepsilon}^{\mathrm{e}} \rangle \cdot \langle \boldsymbol{\sigma}^{\mathrm{t}} \rangle + \langle \boldsymbol{\sigma}^{\mathrm{e}} \cdot \boldsymbol{\varepsilon}_{\theta} \rangle = \langle \boldsymbol{\sigma}^{\mathrm{e}} \cdot \boldsymbol{\varepsilon}_{\theta} \rangle. \end{aligned} \tag{37}$$

Hence,

$$\langle \boldsymbol{\sigma}^{\mathrm{e}} \cdot \boldsymbol{\varepsilon}^{\mathrm{t}} \rangle = \langle \boldsymbol{\sigma}^{\mathrm{e}} \cdot \boldsymbol{\varepsilon}_{\theta} \rangle = \boldsymbol{\Sigma} \cdot \boldsymbol{E}_{\theta}. \tag{38}$$

Applying variation to Eq. $(38)_2$ keeping $\boldsymbol{\varepsilon}_{\theta}$ spatially fixed would result in,

$$\langle \delta\boldsymbol{\sigma}^{\mathrm{e}} \cdot \boldsymbol{\varepsilon}_{\theta} \rangle = \boldsymbol{\Sigma} \cdot \delta\boldsymbol{E}_{\theta}.$$

Equation (34) can now be re-written as,

$$\langle \boldsymbol{\sigma}^{\mathrm{t}} \cdot \delta\mathbb{S}[\boldsymbol{\sigma}^{\mathrm{e}}] \rangle - \langle \delta\boldsymbol{\sigma}^{\mathrm{e}} \cdot \boldsymbol{\varepsilon}_{\theta} \rangle = \langle \boldsymbol{\sigma}^{\mathrm{t}} \cdot \delta\mathbb{S}[\boldsymbol{\sigma}^{\mathrm{e}}] \rangle - \boldsymbol{\Sigma} \cdot \delta\boldsymbol{E}_{\theta} = 0. \tag{39}$$

If the compliance tensor of only $\gamma$-phase is varied by $\delta\mathbb{S}_{\gamma}$ keeping the compliance tensor of other phases and the resulting effective compliance same, then Eq. (39) can be re-written as,

$$c_{\gamma} \langle \boldsymbol{\sigma}^{\mathrm{t}} \otimes \boldsymbol{\sigma}^{\mathrm{e}} \rangle_{\gamma} \cdot \delta\mathbb{S}_{\gamma} = \boldsymbol{\Sigma} \cdot \delta\boldsymbol{E}_{\theta}. \tag{40}$$

Using variational derivative t $\boldsymbol{\Sigma} \cdot \delta\boldsymbol{E}_{\theta}$ can be re-written as,

$$\begin{aligned} \delta\boldsymbol{E}_{\theta} &= \frac{\partial \boldsymbol{E}_{\theta}}{\partial \mathbb{S}_{\gamma}} \cdot \delta\mathbb{S}_{\gamma}, \\ \boldsymbol{\Sigma} \cdot \delta\boldsymbol{E}_{\theta} &= \boldsymbol{\Sigma} \frac{\partial \boldsymbol{E}_{\theta}}{\partial \mathbb{S}_{\gamma}} \cdot \delta\mathbb{S}_{\gamma}. \end{aligned} \tag{41}$$

For any arbitrary choice of $\delta\mathbb{S}_\gamma$, comparing Eq. (41) and (40) the following relation will hold true,

$$\langle \boldsymbol{\sigma}^{\mathrm{t}} \otimes \boldsymbol{\sigma}^{\mathrm{e}} \rangle_\gamma = \frac{1}{c_\gamma} \boldsymbol{\Sigma} \frac{\partial \boldsymbol{E}_\theta}{\partial \mathbb{S}_\gamma}. \tag{42}$$

Similarly, one can show that,

$$\langle \boldsymbol{\sigma}^{\mathrm{e}} \otimes \boldsymbol{\sigma}^{\mathrm{t}} \rangle_\gamma = \frac{1}{c_\gamma} \boldsymbol{\Sigma} \frac{\partial \boldsymbol{E}_\theta}{\partial \mathbb{S}_\gamma}. \tag{43}$$

In index notation, the above expression is written as,

$$\langle \sigma^{\mathrm{e}}_{ij} \sigma^{\mathrm{t}}_{kl} \rangle_\gamma = \frac{1}{c_\gamma} \Sigma_{ab} \frac{\partial E^\theta_{ab}}{\partial S^\gamma_{ijkl}}. \tag{44}$$

For a two-phase composite, $\boldsymbol{E}_\theta$ from Eq. (6) in index notation is,

$$E^\theta_{ij} = \langle \varepsilon_\theta \rangle_{ij} + \left(\bar{S}_{ijab} - \langle S \rangle_{ijab}\right)(\Delta S)^{-1}_{abcd}(\Delta\varepsilon_\theta)_{cd}. \tag{45}$$

In index notation ${\partial \boldsymbol{E}_\theta}/{\partial \mathbb{S}_\gamma}$ is given as,

$$\begin{aligned}\left[\frac{\partial \boldsymbol{E}_\theta}{\partial \mathbb{S}_\gamma}\right]_{ijklmn} = \Bigg( &\left(\frac{\partial \bar{S}_{ijab}}{\partial S^\gamma_{klmn}} - c_\gamma \delta^{8\mathrm{s}}_{ijabklmn}\right)(\Delta S)^{-1}_{abcd} \\ &+ \left(\bar{S}_{ijab} - \langle S \rangle_{ijab}\right)\Delta S^{-1}_{abpq}\delta^{8\mathrm{s}}_{pqrsklmn}\Delta S^{-1}_{rscd}\Bigg)(\Delta\boldsymbol{\varepsilon}_\theta)_{cd}.\end{aligned} \tag{46}$$

In tensor notation,

$$\therefore \frac{\partial \boldsymbol{E}_\theta}{\partial \mathbb{S}_\gamma} = \left( ((\Delta \mathbb{S})^{-1}[\Delta \boldsymbol{\varepsilon}_\theta]) \left( \frac{\partial \overline{\mathbb{S}}}{\partial \mathbb{S}_\gamma} - c_\gamma \boldsymbol{\mathcal{J}}^{\mathrm{s}} \right)^{\mathrm{T_L}} \right. \\ \left. + \Delta \boldsymbol{\varepsilon}_\theta \left( (\overline{\mathbb{S}} - \langle \mathbb{S} \rangle)((\Delta \mathbb{S})^{-1})^{\times 2} \boldsymbol{\mathcal{J}}^{\mathrm{s}} \right)^{\mathrm{T_L}} \right)^{\mathrm{T_R}}. \tag{47}$$

Following to the result obtained in Eq. (15), (27), (42) and (43), the resulting expression for $\langle \boldsymbol{\sigma}^{\otimes 2} \rangle_\gamma$ using Eq. (10) for a linear thermoelastic problem is,

$$\langle \boldsymbol{\sigma}^{\otimes 2} \rangle_\gamma = \frac{1}{c_\gamma} \left( \left( \frac{\partial \overline{\mathbb{S}}}{\partial \mathbb{S}_\gamma} \right)^{\mathrm{T_H}} [\boldsymbol{\Sigma}^{\otimes 2}] + 2 \left( \boldsymbol{\Sigma} \frac{\partial \boldsymbol{E}_\theta}{\partial \mathbb{S}_\gamma} - \frac{\partial W^{\mathrm{t}}}{\partial \mathbb{S}_\gamma} \right) \right). \tag{48}$$

The terms ${}^{\partial \overline{\mathbb{S}}}\!/_{\partial \mathbb{S}_\gamma}$, ${}^{\partial W^{\mathrm{t}}}\!/_{\partial \mathbb{S}_\gamma}$, and ${}^{\partial \boldsymbol{E}_\theta}\!/_{\partial \mathbb{S}_\gamma}$ obtained for a two-phase composite in Eq. (16), (29) and (47) respectively can be used along with Eq. (18) to compute the second moments. The anisotropic derivative of the Hill's polarization tensor in Eq. (19) remains unknown which is evaluated in the next section.

## 4 Derivatives of the Polarization Tensor with Reference Stiffness Tensor ($\partial \mathbb{P}_{\mathbf{o}} / \partial \mathbb{C}_{\mathbf{o}}$)

Consider an RVE of a heterogenous material comprising of two-phases in which multiple non-overlapping and identical ellipsoidal (prolate spheroid) shaped fibers are uniformly distributed in a matrix (see Fig. 2). The aspect ratio of the ellipsoidal fiber is defined as $a = l/d$, where $l$ and $d$ represents the major and the minor diameter of the fiber, respectively. The major diameter of the ellipsoid is aligned along the $\boldsymbol{e}_3$-direction (indicated in Fig. 2 with right-handed co-ordinate system). The spatial arrangement of the ellipsoidal inhomogeneities (described by the two-point correlation function), is assumed to follow a ellipsoidal distribution that is characterized by the shape of the inhomogeneity itself [11, 24, 25]. Let the compliant phase be the matrix (assigned as $\gamma = 1$), characterized by an elastic stiffness tensor $\mathbb{C}_1$ or $\mathbb{C}_\mathrm{m}$ whereas the stiffer reinforcements are characterized by same elastic stiffness tensor $\mathbb{C}_2$ or $\mathbb{C}_\mathrm{f}$. The volume fraction of the matrix is indicated as $c_1$ and the volume fraction of fiber phase is $c_2 = 1 - c_1$. The effective stiffness of the heterogenous material can be found using Eq. (4) and (5). Using the definition of the polarization tensor as defined in Eq. (5) it can be re-written in index notation as,

$$
\begin{aligned}
&P^{\mathrm{o}}_{ijkl} = \frac{1}{4\pi a^2}\int_S c(\theta) H_{ijkl}(\mathbb{C}_\mathrm{o}, \boldsymbol{n})\mathrm{d}S,\\
&c(\theta,\varphi) = \Big(1 + \big((a^{-2}-1)\sin^2\theta\big)\Big)^{-\frac{3}{2}},\\
&K_{lj} = C^{\mathrm{o}}_{lmjn}N_{mn},\\
&H_{ijkl} = \frac{1}{4}(N_{ik}(K_{lj})^{-1} + N_{jk}(K_{li})^{-1} + N_{il}(K_{kj})^{-1} + N_{jl}(K_{ki})^{-1}).
\end{aligned}
\tag{49}
$$

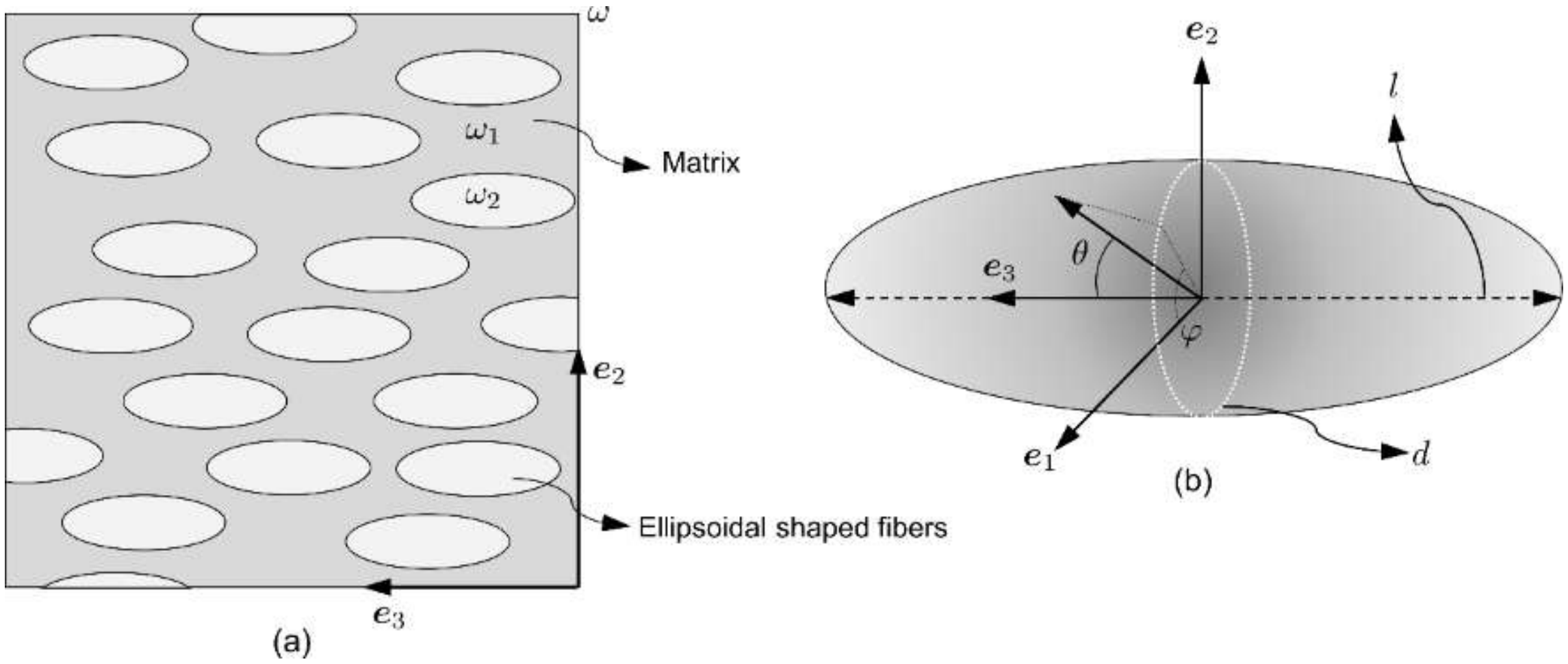

**Fig. 2.** (a) Representative volume element, $\omega$ with random distribution of non-overlapping ellipsoids and oriented along $\boldsymbol{e}_3$-direction in a matrix. (b) Ellipsoid with major and minor diameter $l$ and $d$, respectively.

The derivatives $\partial\mathbb{P}_\mathrm{o}/\partial\mathbb{C}_\mathrm{o}$ is then given as,

$$\frac{\partial\mathbb{P}_\mathrm{o}}{\partial\mathbb{C}_\mathrm{o}} = \frac{1}{4\pi a^2}\int_S c(\theta)\frac{\partial\mathbb{H}(\mathbb{C}_\mathrm{o},\boldsymbol{n})}{\partial\mathbb{C}_\mathrm{o}}\mathrm{d}S,$$

$$\frac{\partial\mathbb{H}(\mathbb{C}_\mathrm{o},\boldsymbol{n})}{\partial\mathbb{C}_\mathrm{o}} = \mathbb{I}^\mathrm{s}\left(\frac{\partial(\boldsymbol{N}\times\boldsymbol{K}^{-1})}{\partial\mathbb{C}_\mathrm{o}}\right)\mathbb{I}^\mathrm{s}, \qquad \frac{\partial(\boldsymbol{N}\times\boldsymbol{K}^{-1})}{\partial\mathbb{C}_\mathrm{o}} = -\boldsymbol{N}\times\boldsymbol{K}^{-1}\times\boldsymbol{K}^{-1}\times\boldsymbol{N}.$$

In index notation,

$$\frac{\partial P^\mathrm{o}_{ijkl}}{\partial C^\mathrm{o}_{opqr}} = \frac{1}{4\pi a^2}\int_S c(\theta)\frac{\partial H_{ijkl}}{\partial C^\mathrm{o}_{opqr}}\mathrm{d}S,$$

$$\frac{\partial H_{ijkl}}{\partial C^\mathrm{o}_{opqr}} = \frac{1}{4}\left(N_{ik}\frac{\partial {K_{lj}}^{-1}}{\partial C^\mathrm{o}_{opqr}} + N_{jk}\frac{\partial {K_{li}}^{-1}}{\partial C^\mathrm{o}_{opqr}} + N_{il}\frac{\partial {K_{kj}}^{-1}}{\partial C^\mathrm{o}_{opqr}} + N_{jl}\frac{\partial {K_{ki}}^{-1}}{\partial C^\mathrm{o}_{opqr}}\right), \tag{50}$$

$$\frac{\partial {K_{lj}}^{-1}}{\partial C^\mathrm{o}_{opqr}} = -K^{-1}_{lk} I^{8s}_{kmsnopqr} N_{mn} K^{-1}_{sj}.$$

Phases of heterogenous materials used in commercial applications, such as polymer composites found in automotive applications, are typically isotropic. For such class of composite materials better estimates are made by choosing an isotropic reference stiffness ($\mathbb{C}_\mathrm{o} = \mathbb{C}_\mathrm{m}$). The isotropic fourth-order stiffness tensor has two eigenvalues i.e., the bulk and shear moduli. The corresponding acoustic tensor is composed of an isotropic contribution arising from the shear modulus, together with a directional contribution associated with the bulk response along the propagation direction. A closed-form inverse exists because this structure is essentially a rank-one modification of a scaled identity, which makes the Sherman–Morrison formula directly applicable. Hence, closed form of expressions of derivatives and the integrals in Eq. (50) can be found analytically. The analytical solution of the eighth order tensor $\partial\mathbb{P}_\mathrm{o}/\partial\mathbb{C}_\mathrm{o}$ for isotropic case is provided in Ref. [1]. From an implementation perspective, analytical expressions enable quick evaluation of field fluctuations. In contrast, for anisotropic constituents and reinforcements with non-ellipsoidal shapes, numerical integration methods are essential to solve Eq. (50).

## 5 Sampling and Statistics of Local Fields

First and second moments of the random stress field quantity are obtained using Eq. (7) and (48). The uniform spatial distribution of inhomogeneities in the microstructure influences the field quantity of interest at any matrix point due to inhomogeneity-matrix interactions and spatial arrangements. Following the central limit theorem, as the RVE size increases, the combined effects of these random stress contributions approach a Gaussian distribution. Consequently, it is assumed that the tensorial local field variables, as well as their linear combinations within each phase, tends towards a Gaussian distribution. In this case, sampling of the local field variables, e.g., stress tensor, based on the first and second moments of the stress tensor is possible. This is given as,

$$\boldsymbol{\sigma} = \langle \boldsymbol{\sigma} \rangle_\gamma + \mathbb{L}[\boldsymbol{\chi}], \tag{51}$$

where $\boldsymbol{\sigma}$ is the sampled stress tensor, $\mathbb{L}$ is the Cholesky decomposition of the positive semi-definite variance tensor $\mathbb{K}^\sigma$ such that $\mathbb{K}^\sigma = \mathbb{L}\mathbb{L}^{\mathrm{T_H}}$ and $\boldsymbol{\chi}$ is a second order tensor whose components are independent normal random variables with $[\langle \boldsymbol{\chi} \rangle]_{ij} = 0$ and $[\langle \boldsymbol{\chi}^2 \rangle]_{ij} = 1$. From the sampled stress tensor, invariants of the tensorial field quantities are sampled, such as the equivalent or von-Mises stress $\left(\sigma_{\mathrm{eq}} = \sqrt{\frac{3}{2}\boldsymbol{\sigma}' \cdot \boldsymbol{\sigma}'}\right)$.

The Gaussian probability density of a scalar quantity, e.g. $(\vartheta)$ with normal distribution $\mathcal{N}(\langle \vartheta \rangle, K^\vartheta)$ or any second order tensor (muti-variate), e.g. $(\boldsymbol{\vartheta})$ with distribution $\mathcal{N}(\langle \boldsymbol{\vartheta} \rangle, \mathbb{K}^{\boldsymbol{\vartheta}})$ can be evaluated using,

$$\begin{aligned} p(\vartheta) &= \frac{1}{\sqrt{2\pi\sigma}} \exp\left(-\frac{1}{2}(K^\vartheta)^{-1}(\vartheta - \langle \vartheta \rangle)^2\right), \\ p(\boldsymbol{\vartheta}) &= \frac{1}{\sqrt{(2\pi)^6 \det(\mathbb{K}^{\boldsymbol{\vartheta}}_\gamma)}} \exp\left(-\frac{1}{2}(\mathbb{K}^{\boldsymbol{\vartheta}}_\gamma)^{-1} \cdot (\boldsymbol{\vartheta} - \langle \boldsymbol{\vartheta} \rangle_\gamma)^{\otimes 2}\right). \end{aligned} \tag{52}$$

## 6 Field Statistics from Full-Field Homogenization

The exact statistical distribution of local fields is obtained through full-field analysis, which is used to validate the outcomes discussed in Section 3 and 4. In this study, full-field analysis of microstructures is performed using finite elements for microstructures of unidirectional fiber reinforced polymer composites and particulate composites. A uniform distribution of non-overlapping long fibers with a unidirectional (UD) arrangement within a matrix is considered for long fiber reinforced polymer (FRP) composite microstructures whereas spherical shaped glass particles ($a = 1$) are considered for microstructures of particulate composites. To demonstrate the outcome of the derived equations, volume fraction of reinforcements are restricted below $c_{\mathrm{f}} < 0.25$ as noticed in automotive applications [26]. In view of this, a random sequence adsorption algorithm is sufficient to generate artificial microstructures for full-field analysis. This approach is implemented in an in-house developed code to generate random position vectors of inhomogeneities considering the constraints of periodicity and non-overlap. The geometrical model generation of the artificial microstructures using these position vectors and discretizing with 10-node quadratic tetrahedron elements are seamlessly done using NETGEN [27]. An in-house code is used to apply periodic boundary condition on the free surfaces of the RVE [28]. The traction and displacements are considered to be continuous across the interface of the matrix and inhomogeneities. Numerical analysis of the microstructures is performed in a commercial FE solver ABAQUS.

In this work, glass material reinforced in unsaturated polyester–polyurethane hybrid (UPPH) polymer is considered (elastic contrast of 21.5). Experimentally characterized properties of the phases are listed in Table 2. In a typical composite manufacturing process like compression molding, after curing process of thermoset matrix the composite part cools from temperature of 140 °C to room temperature conditions [26]. At the end of the cooling process, residual stress builds up due to differential expansion of glass fibers and the UPPH matrix. A stress-free strain condition is defined by setting $\boldsymbol{E} = \boldsymbol{E}_{\boldsymbol{\theta}}$ to predict the statistical distribution of localized residual stress generated purely due to differential thermal expansion.

**Table 2.** Linear thermoelastic properties of the phases in heterogenous materials. [29, 30]

| Property | UPPH Matrix | Glass Fiber |
|---|---|---|
| Elastic Modulus (GPa) | 3.4 | 73 |
| Poisson's Ratio | 0.385 | 0.22 |
| Coefficient of Thermal Expansion ($K^{-1}$) | 7.4e-5 | 0.5e-5 |

## 7 Validations: Field Statistics and Residual Stress

### 7.1 UDFRP Composites

The term $\partial \mathbb{P}_o / \partial \mathbb{C}_o$ in Eq. (50) is evaluated for inhomogeneities with $a = 1000$ (cylindrical shaped long fibers) and $c_2 = 0.25$ to find the first and second moments of the stress field in a linear thermoelastic homogenization problem. Full-field simulations for long FRP microstructures with 20 cylindrical shaped particles are carried out to extract the statistics of the local residual stress fields.

Fig. 3 (a) and (b) shows the statistical distribution of the normal and shear components of the residual stress tensor in the matrix domain generated purely due to CTE mismatch of the phases after cooling of the composite. The average normal and shear stresses are 6.48 MPa and 0 MPa, respectively, yet both show localized peaks in the range of ±40 MPa. These extremes, rather than the averages, are critical in initiating progressive damage and driving composite failure under loading. Further, the statistical estimates obtained by sampling the stress data from the computed first and second moments using the mean-field solution is comparable to the full-field (exact) solution. The statistical distribution doesn't exactly match as the histogram contains the higher order statistical moments. Thereby mean-field offers a major advantage that the statistical estimates of the local field can be computed instantly and at a fraction of the cost of computationally expensive full-field simulations.

Fig. 4 shows the equivalent stress computed from the sampled stress data using the first and second moments obtained from the mean-field solution. It can be observed that though the components of stress exhibit a Gaussian distribution, the equivalent residual stress tend to follow more towards a Weibull distribution under stress-free strain cooling of long fiber reinforced composite. The peak stress values are more concentrated in the boundary region of the fibers or matrix region lying between two closely positioned fibers. The other major of the volume region experiences lower peak of the residual stress.

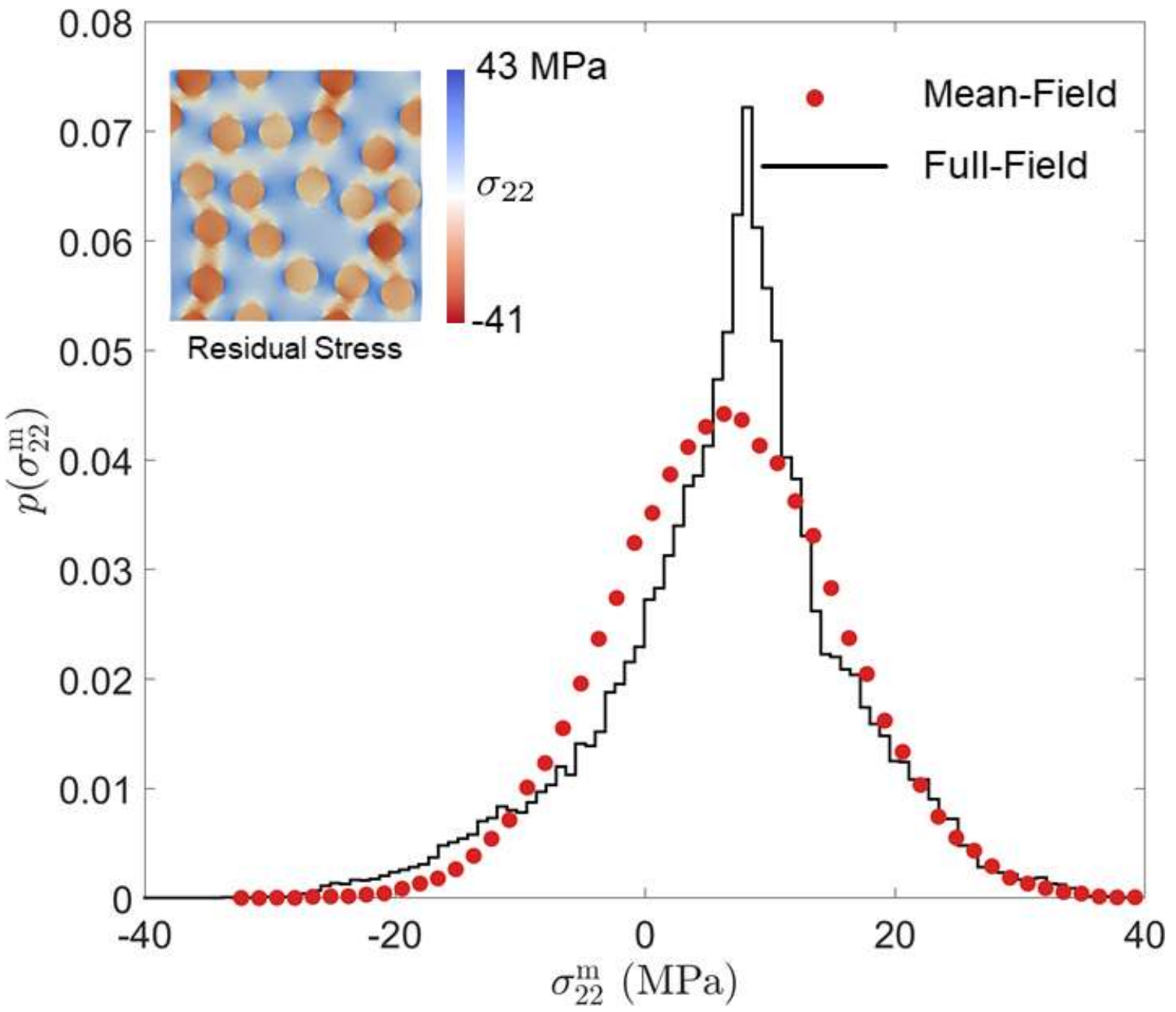

(a)

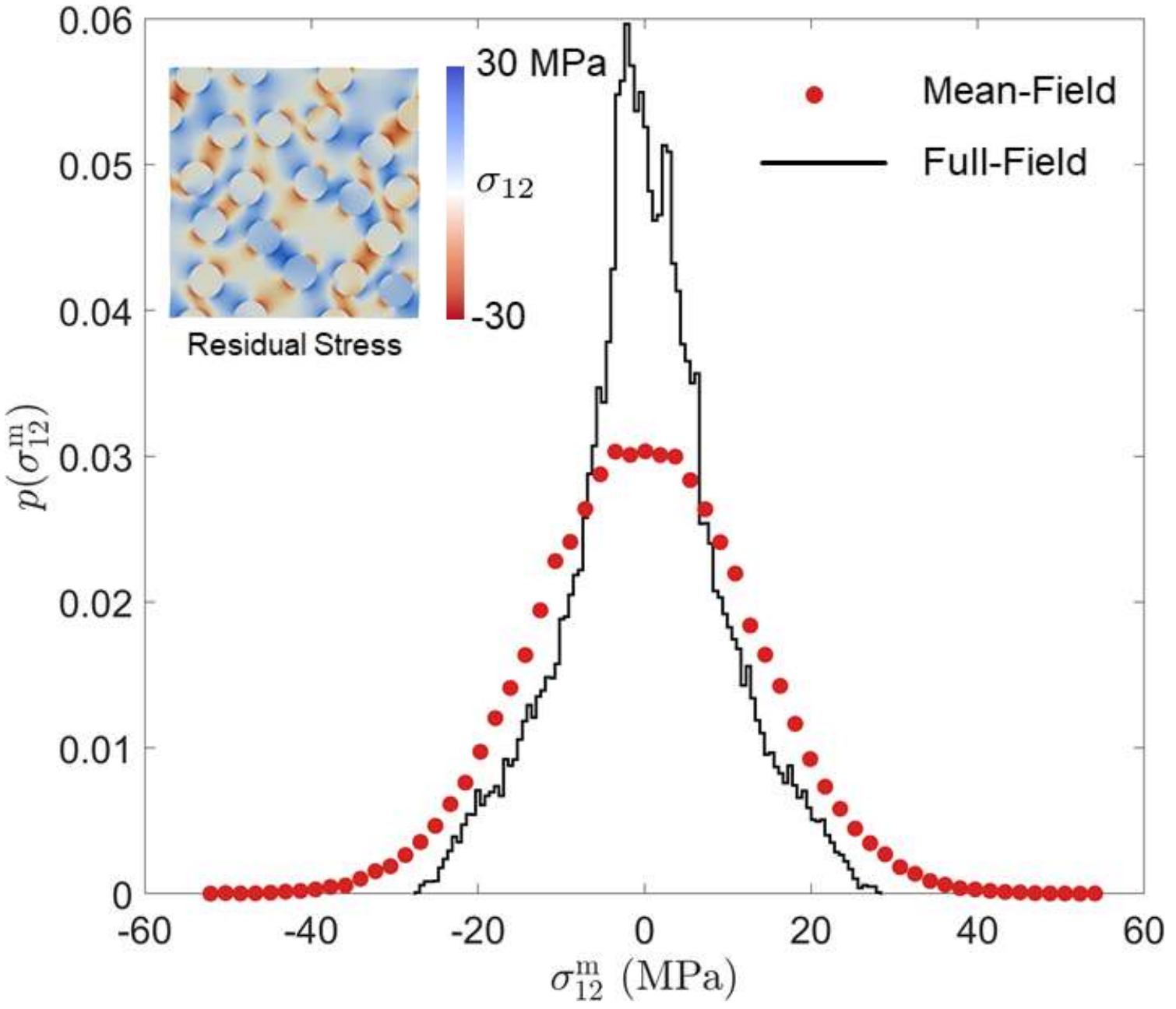

(b)

**Fig. 3.** Statistical distribution of localized residual stress in the matrix domain of long fiber reinforced polymer composite (a) normal (22) component (b) shear (12) component of the stress tensor. Figure inset shows the exact distribution of localized stress in matrix as well as fiber domain. The fibers, with a volume fraction of 25%, are aligned along the $\boldsymbol{e}_3$ direction (out of paper plane).

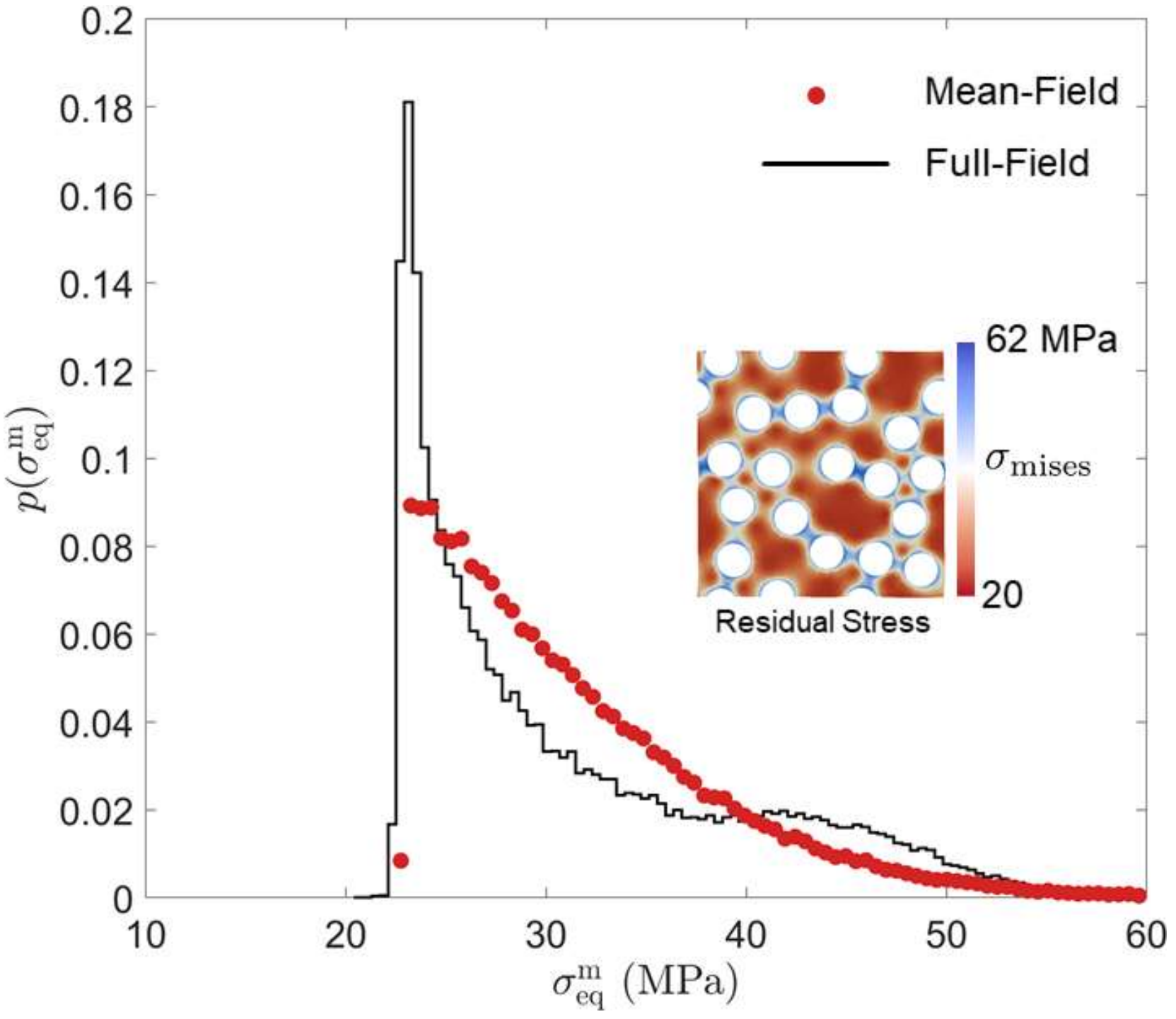

**Fig. 4.** Statistical distribution of localized equivalent stress in the matrix domain of long fiber reinforced polymer composite with 25% volume fraction of fibers along $\boldsymbol{e}_3$ direction (out of the plane of paper). Results are compared from mean-field and full-field (exact). Figure inset shows the exact distribution of localized stress in the matrix domain.

### 7.2 Particulate Composites

Fig. 5 shows the statistical distribution of components of residual stress tensor purely due to CTE mismatch of the phases after cooling of the particulate composite ($a = 1$, $c_2 = 0.25$). The exact distribution of the components of the stress tensor can be approximated by a Gaussian distribution. The solution obtained from the mean-field is comparable to the full-field (exact) solution though not exactly matching as the histogram.

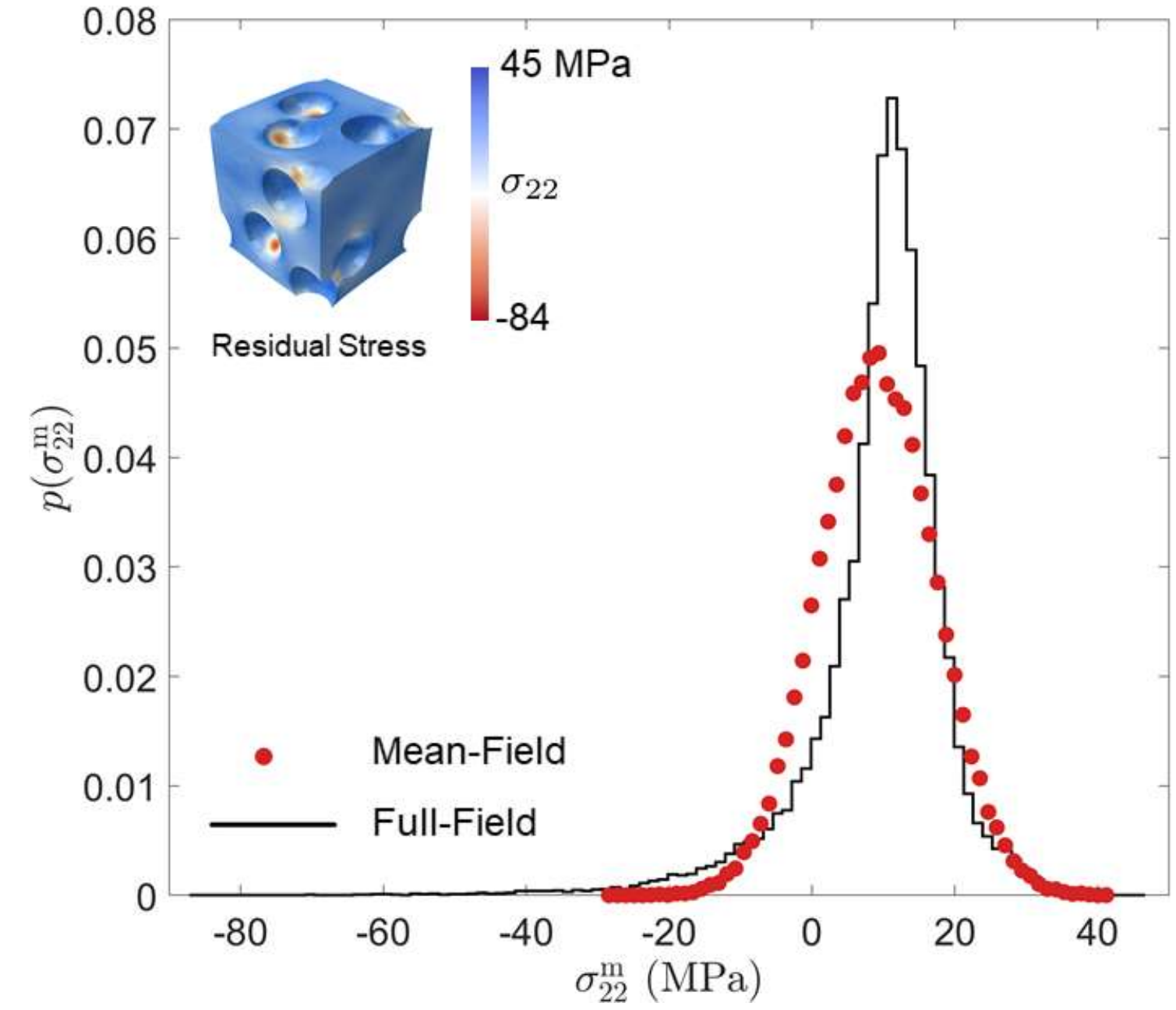

(a)

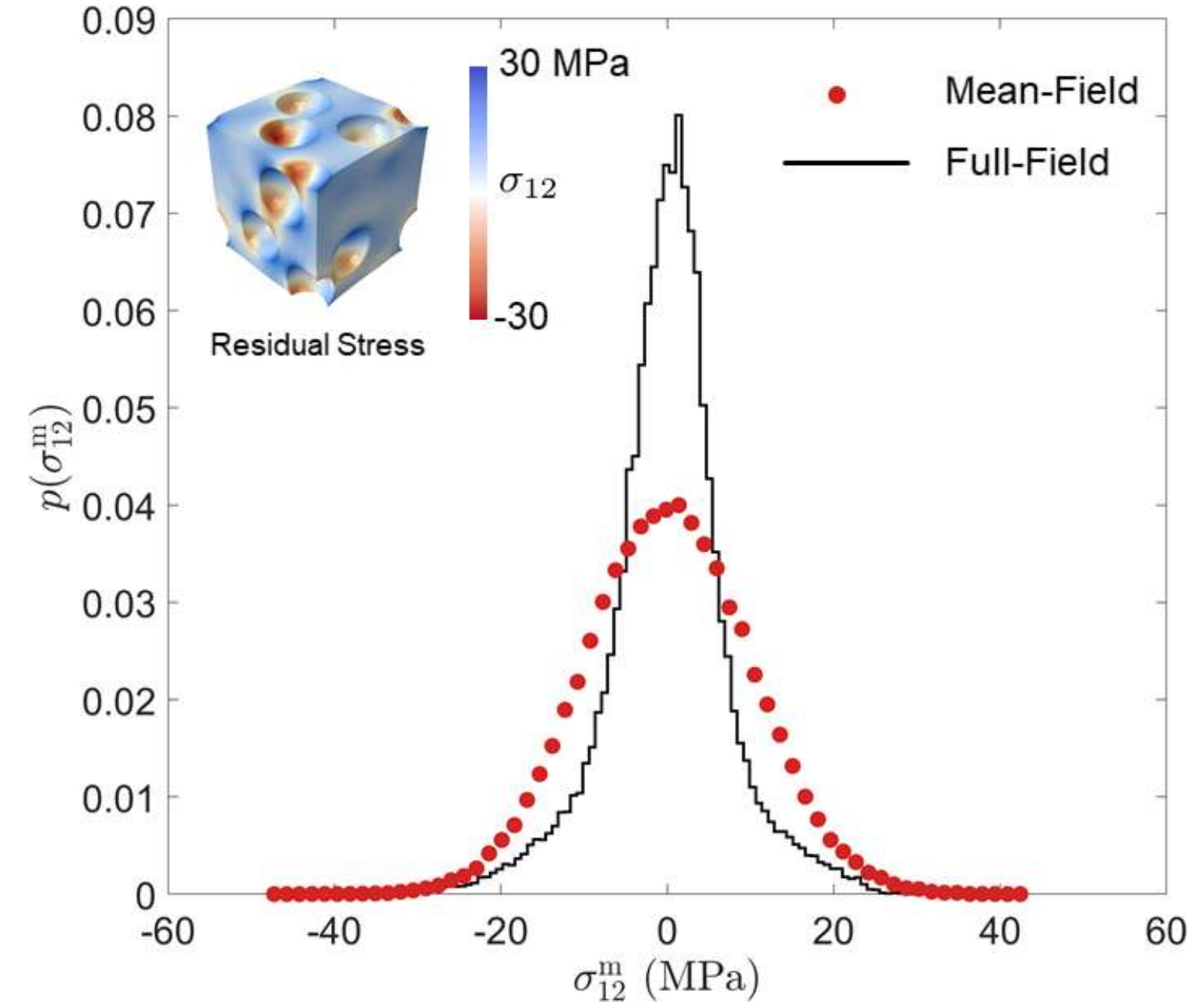

(b)

**Fig. 5.** Statistical distribution of localized residual stress in continuous long fiber reinforced polymer composite (a) normal (22) component (b) shear (12) component of the stress tensor.

Fig. 6 shows the equivalent stress computed from the sampled stress tensor. From the exact solution it can be observed that the components of stress can be approximated by Gaussian distribution from mean-field under stress-free strain cooling of particulate composite. However, the actual distribution of von Mises stress distribution indicates Weibull-type which is not captured by the Gaussian approximation of the distribution assumed from mean-field. The peak stress values are more concentrated in the boundary region of the spherical inhomogeneities.

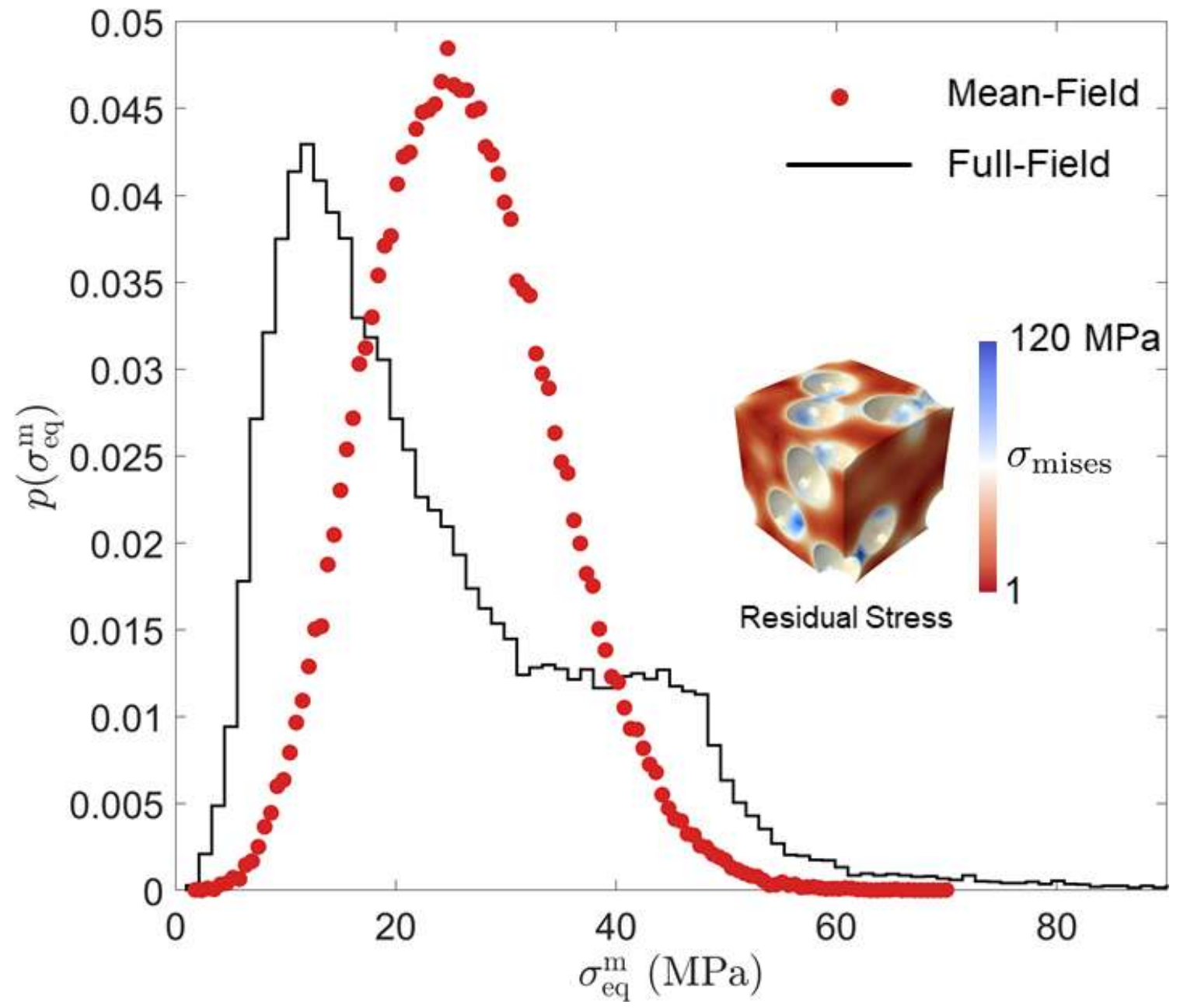

**Fig. 6.** Statistical distribution of localized von-Mises stress in the matrix domain of the particulate composite with $c_2 = 0.25$.

## 8 Conclusions

Closed-form relations for computing first and second moments of local stress fields are derived within the mean-field homogenization framework, relying on derivatives of Hill's polarization tensor. Later the solution is validated with full-field finite element simulations of random microstructures of UD FRP and particulate composites. These expressions are shown to provide an efficient means of estimating field fluctuations without resorting to computationally intensive simulations.

Due to the availability of the first and second moments, a gaussian distribution of stress components is assumed from mean-field estimates using central limit theorem applied to random composites. Validation against full-field homogenization of fibrous and particulate composites subjected to stress-free strain cooling revealed that mean-field can capture the statistical trends of residual stress distributions, both for individual tensor components and their invariants. However, from full-field, the actual distribution of stress components is non-gaussian type whereas the equivalent residual stress showed Weibull-like distributions.

Overall, mean-field provides a computationally inexpensive yet reasonably accurate tool for estimating local field statistics in thermoelastic composites. The approach offers direct applicability in the prediction of process-induced residual stresses and in assessing the statistical variability relevant for failure initiation in engineering composites.

Acknowledgments

The author would like to acknowledge the support of Institute Startup Grant of IIT Guwahati.

**References**

1. Pallicity, T.D., Krause, M., Böhlke, T.: Local field statistics in linear elastic unidirectional fibrous composites. Int. J. Solids Struct. 315, 113343 (2025). https://doi.org/10.1016/j.ijsolstr.2025.113343

2. Lahellec, N., Suquet, P.: Effective behavior of linear viscoelastic composites: A time-integration approach. Int. J. Solids Struct. 44, 507–529 (2007). https://doi.org/10.1016/j.ijsolstr.2006.04.038

3. Badulescu, C., Lahellec, N., Suquet, P.: Field statistics in linear viscoelastic composites and polycrystals. Eur. J. Mech. A/Solids. 49, 329–344 (2015).

https://doi.org/10.1016/j.euromechsol.2014.07.012

4. Pallicity, T.D., Böhlke, T.: Effective viscoelastic behavior of polymer composites with regular periodic microstructures. Int. J. Solids Struct. 216, 167–181 (2021). https://doi.org/10.1016/j.ijsolstr.2021.01.016

5. Kowalczyk-Gajewska, K., Berbenni, S., Mercier, S.: An additive Mori–Tanaka scheme for elastic–viscoplastic composites based on a modified tangent linearization. Mech. Mater. 200, 105191 (2024). https://doi.org/10.1016/j.mechmat.2024.105191

6. Bobeth, M., Diener, G.: Field fluctuations in multicomponent mixtures. J. Mech. Phys. Solids. 34, 1–17 (1986). https://doi.org/10.1016/0022-5096(86)90002-5

7. Bobeth, M., Diener, G.: Static elastic and thermoelastic field fluctuations in multiphase composites. J. Mech. Phys. Solids. 35, 137–149 (1987). https://doi.org/10.1016/0022-5096(87)90033-0

8. Kreher, Wolfgang; Pompe, W.: Internal stresses in heterogeneous solids. John Wiley & Sons Australia, Limited (1989)

9. Brenner, R., Castelnau, O., Badea, L.: Mechanical Field Fluctuations in Polycrystals Estimated by Homogenization Techniques. Source Proc. Math. Phys. Eng. Sci. 460, 3589–3612 (2004). https://doi.org/10.1098/rspa.2004.1278

10. Willot, F., Brenner, R., Trumel, H.: Elastostatic field distributions in polycrystals and cracked media. Philos. Mag. 100, 661–687 (2020). https://doi.org/10.1080/14786435.2019.1699669

11. Das, S., Ponte Castañeda, P.: Field statistics in linearized elastic and viscous composites and polycrystals. Int. J. Solids Struct. 224, 111030 (2021). https://doi.org/10.1016/j.ijsolstr.2021.03.017

12. Wismans, M., van Breemen, L.C.A., Govaert, L.E., Engels, T.A.P.: The Effect of Thermal Residual Stress on the Stress State in a Short-Fiber Reinforced Thermoplastic. J. Mater. Eng. Perform. 33, 4160–4169 (2024). https://doi.org/10.1007/s11665-024-09277-x

13. Anvari, A.: Thermomechanical Fatigue of Unidirectional Carbon Fiber/Epoxy Composite in Space. J. Eng. 2020, 1–5 (2020). https://doi.org/10.1155/2020/9702957

14. Zheng, Q.-S., Spencer, A.J.M.: On the canonical representations for kronecker powers of orthogonal tensors with application to material symmetry problems. Int. J. Eng. Sci. 31, 617–635 (1993). https://doi.org/10.1016/0020-7225(93)90054-X

15. Torquato, S.: Random Heterogeneous Materials. Springer New York, New York, NY (2002)

16. Fokin, A.G.: Solution of statistical problems in elasticity theory in the singular approximation. J. Appl. Mech. Tech. Phys. 13, 85–89 (1974). https://doi.org/10.1007/BF00852360

17. Jöchen, K.: Homogenization of the Linear and Non-linear Mechanical Behavior of Polycrystals, (2013)

18. Lobos Fernández, M.: Homogenization and materials design of mechanical properties of textured materials based on zeroth-, first- and second-order bounds of linear behavior, https://publikationen.bibliothek.kit.edu/1000080683, (2018)

19. Krause, M., Pallicity, T.D., Böhlke, T.: Exact second moments of strain for composites with isotropic phases. Eur. J. Mech. - A/Solids. 97, 104806 (2023). https://doi.org/10.1016/j.euromechsol.2022.104806

20. Buryachenko, V.A.: Micromehcanics of Heterogenous Materials. Springer US,

Boston, MA (2007)

21. Rosen, B.W., Hashin, Z.: Effective thermal expansion coefficients and specific heats of composite materials. Int. J. Eng. Sci. 8, 157–173 (1970). https://doi.org/10.1016/0020-7225(70)90066-2

22. Kreher, W.: Internal Stresses and Relations between Effective Thermoelastic Properties of Stochastic Solids – Some Exact Solutions. ZAMM - J. Appl. Math. Mech. / Zeitschrift für Angew. Math. und Mech. 68, 147–154 (1988). https://doi.org/10.1002/zamm.19880680311

23. Laws, N.: On the thermostatics of composite materials. J. Mech. Phys. Solids. 21, 9–17 (1973). https://doi.org/10.1016/0022-5096(73)90027-6

24. Hu, G.K., Weng, G.J.: Connections between the double-inclusion model and the Ponte Castaneda-Willis, Mori-Tanaka, and Kuster-Toksoz models. Mech. Mater. 32, 495–503 (2000). https://doi.org/10.1016/S0167-6636(00)00015-6

25. Castañeda, P.P., Willis, J.R.: The effect of spatial distribution on the effective behavior of composite materials and cracked media. J. Mech. Phys. Solids. 43, 1919–1951 (1995). https://doi.org/10.1016/0022-5096(95)00058-Q

26. Görthofer, J., Meyer, N., Pallicity, T.D., Schöttl, L., Trauth, A., Schemmann, M., Hohberg, M., Pinter, P., Elsner, P., Henning, F., Hrymak, A., Seelig, T., Weidenmann, K., Kärger, L., Böhlke, T.: Virtual process chain of sheet molding compound: Development, validation and perspectives. Compos. Part B Eng. 169, (2019). https://doi.org/10.1016/j.compositesb.2019.04.001

27. Schöberl, J.: NETGEN An advancing front 2D/3D-mesh generator based on abstract rules. Comput. Vis. Sci. 1, 41–52 (1997). https://doi.org/10.1007/s007910050004

28. Albiez, J., Erdle, H., Weygand, D., Böhlke, T.: A gradient plasticity creep model

accounting for slip transfer/activation at interfaces evaluated for the intermetallic NiAl-9Mo. Int. J. Plast. 113, 291–311 (2019). https://doi.org/10.1016/j.ijplas.2018.10.006

29. Trauth, A., Weidenmann, K.A.: Continuous-discontinuous sheet moulding compounds – Effect of hybridisation on mechanical material properties. Compos. Struct. 202, 1087–1098 (2018). https://doi.org/10.1016/j.compstruct.2018.05.048

30. Kehrer, L.: Thermomechanical mean-field modeling and experimental characterization of long fiber-reinforced sheet molding compound composites, (2019)